\documentclass[aps,prb,twocolumn,superscriptaddress]{revtex4-1}

\usepackage{graphicx}
\usepackage{amsmath}
\usepackage{amssymb}
\usepackage{float}
\usepackage{placeins}
\usepackage{verbatim}
\usepackage{natbib}
\usepackage[utf8]{inputenc}
\usepackage[T1]{fontenc}

\begin{document}


\title{Nanoscale X-Ray Imaging of Spin Dynamics in Yttrium Iron Garnet}

\author{J. F\"orster}
\affiliation{Max-Planck-Institute for Intelligent Systems, Stuttgart, Germany}
\author{S. Wintz}
\affiliation{Paul Scherrer Institute, Villigen, Switzerland}
\affiliation{Helmholtz-Zentrum Dresden-Rossendorf, Germany}
\author{J. Bailey}
\affiliation{Paul Scherrer Institute, Villigen, Switzerland}
\affiliation{École polytechnique fédérale de Lausanne (EPFL), Lausanne, Switzerland}
\author{S. Finizio}
\affiliation{Paul Scherrer Institute, Villigen, Switzerland}
\author{E. Josten}
\affiliation{Helmhotz-Zentrum Dresden-Rossendorf, Germany}
\affiliation{Ernst Ruska-Centrum f\"ur Mikroskopie und Spektroskopie mit Elektronen, Forschungszentrum J\"ulich GmbH, J\"ulich, Germany}
\author{D. Meertens}
\affiliation{Ernst Ruska-Centrum f\"ur Mikroskopie und Spektroskopie mit Elektronen, Forschungszentrum J\"ulich GmbH, J\"ulich, Germany}
\author{C. Dubs}
\affiliation{INNOVENT e.V. Technologieentwicklung Jena, Germany}
\author{D. A. Bozhko}
\affiliation{Technische Universit\"at Kaiserslautern, Germany}
\affiliation{University of Glasgow, United Kingdom}
\author{H. Stoll}
\affiliation{Max-Planck-Institute for Intelligent Systems, Stuttgart, Germany}
\affiliation{Institut f\"ur Physik, Johannes Gutenberg-Universit\"at Mainz, Germany}
\author{G. Dieterle}
\author{N. Tr\"ager}
\affiliation{Max-Planck-Institute for Intelligent Systems, Stuttgart, Germany}
\author{J. Raabe}
\affiliation{Paul Scherrer Institute, Villigen, Switzerland}
\author{A. N. Slavin}
\affiliation{Oakland University, Rochester, USA}
\author{M. Weigand}
\affiliation{Max-Planck-Institute for Intelligent Systems, Stuttgart, Germany}
\affiliation{Helmholtz-Zentrum Berlin, Germany}
\author{J. Gr\"afe}
\author{G. Sch\"utz}
\affiliation{Max-Planck-Institute for Intelligent Systems, Stuttgart, Germany}


\date{\today}

\begin{abstract}
Time-resolved scanning transmission x-ray microscopy (TR-STXM) has been used for the direct imaging of spin wave dynamics in thin film yttrium iron garnet (YIG) with spatial resolution in the sub 100 nm range. Application of this x-ray transmission technique to single crystalline garnet films was achieved by extracting a lamella (13x5x0.185 $\mathrm{\mu m^3}$) of liquid phase epitaxy grown YIG thin film out of a gadolinium gallium garnet substrate. Spin waves in the sample were measured along the Damon-Eshbach and backward volume directions of propagation at gigahertz frequencies and with wavelengths in a range between 100~nm and 10~$\mathrm{\mu}$m. The results were compared to theoretical models. Here, the widely used approximate dispersion equation for dipole-exchange spin waves proved to be insufficient for describing the observed Damon-Eshbach type modes. For achieving an accurate description, we made use of the full analytical theory taking mode-hybridization effects into account. 
\end{abstract}


\maketitle

\section{Introduction \label{Intro}}

\begin{figure}[h]
	\includegraphics[width=0.4\textwidth]{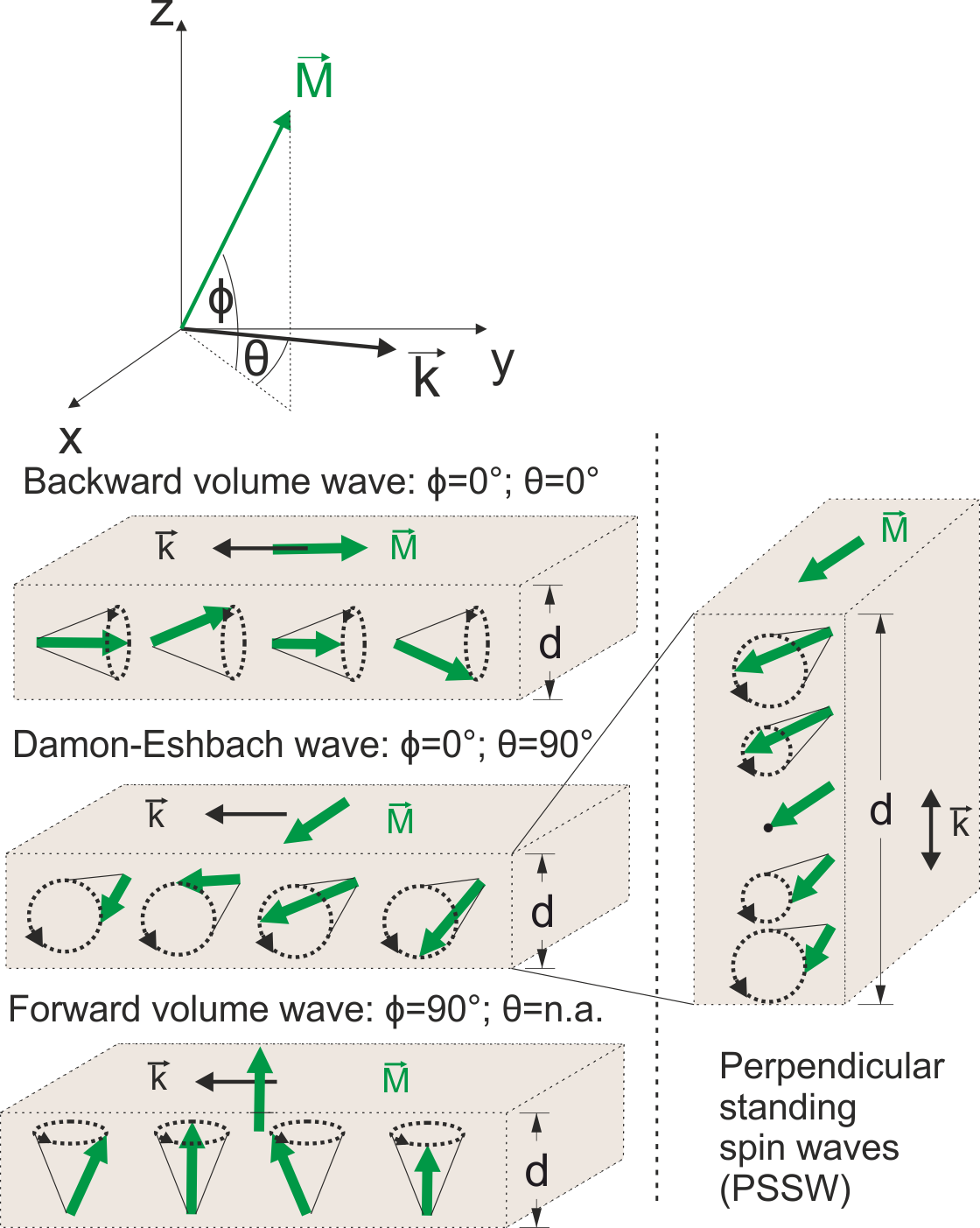}
	\caption{Overview of the basic spin wave mode geometries in a magnetic thin film of thickness $d$. Green arrows symbolize the magnetization vector $\textbf{M}$, while the black ones show the wave vector $\textbf{k}$. \label{figProfile}}
\end{figure}

Spin waves are collective magnetic excitations in \mbox{ferro-,} ferri- and antiferromagnetic materials and an active research area in the field of magnetism. Recently, it was demonstrated that their quanta, magnons, show specific fundamentals of bosonic behaviour such as Bose Einstein condensation and super-fluidity \cite{Demokritov06, Bozhko16}. And, even black hole scenarios have been predicted to occur in magnon gases \cite{Roldan_Molina17}. Besides the fundamental impact of this topic, there has also been increasing interest in potential applications of spin waves as information carriers. This has led to the emergence of the field of magnonics. Compared to electromagnetic waves, spin-wave wavelengths are smaller by several orders of magnitude, which fits perfectly to the lateral dimensions of $10\ nm\ - 1\ \mathrm{\mu m}$ achievable by modern nanotechnology. Spin waves excellently cover the Gigahertz-regime of frequencies, which is common in today’s communications devices, allowing their creation and detection via well-developed microwave techniques. Furthermore, and in contrast to conventional electronics, spin waves can carry information without power dissipating charge currents. Therefore spin waves are actively discussed as high-speed and short-wavelength information carriers for novel spintronic/magnonic devices \cite{Serga10, Chumak15}. Magnetic thin film systems exhibit three basic geometries for lateral spin wave propagation in their spectrum (\textit{c.f.} figure \ref{figProfile}). For in-plane magnetized films there is the backward volume (BV) geometry with waves propagating along the equilibrium magnetization direction, as well as the Damon-Eshbach (DE) geometry, in which the waves propagate perpendicular to it. Forward volume waves occur in films magnetized out-of-plane propagating isotropically in any direction in the film plane \cite{Serga10, Stancil09, Damon65, Liu15, Klingler15_2}. Finally, in addition to the fundamental modes with a quasi-uniform amplitude profile over the film thickness, all three geometries possess higher order thickness modes with amplitude profiles in the form of perpendicular standing spin waves (PSSW) between the two film surfaces \cite{Kalinikos86}. The relevant energy contributions that determine the dispersion relations $f(k=2\pi/\lambda)$ of the spin waves in these geometries are the magnetostatic and exchange interactions, which dictate the long and short wavelength regimes respectively \cite{Kalinikos86}.

The insulating ferrimagnet yttrium iron garnet (YIG) is one of the most prominent and extensively studied materials in the field of magnonics due to its exceptionally low magnetic damping and high spin wave propagation length, making it ideal as a model system \cite{Bozhko17, Kreil18, Serga14} and for possible applications \cite{Serga10, Damon65, Kaak99, Baek04, Fischer17}. Studies become sparse, however, for wavelengths and spatial features below 250~nm, despite the importance of this regime for potential nanoscale spintronic devices and the open questions it holds. Factors such as surface effects, crystal defects, grain sizes or spin diffusion become more influential on this scale \cite{Nembach13, Adur14} and can change spin wave behaviour compared to the well-studied microscale. For example, an increase in spin wave damping as well as the emergence of frequency dependent damping \cite{Nembach13, Adur14} are expected at the nanoscale. The main reason for this region being less well studied lies in its experimental accessibility. Direct imaging of spin wave dynamics is conventionally performed by optical techniques like Kerr microscopy \cite{Talalaevskij17, Park02} and Brillouin light scattering\cite{Serga10, Chumak15, Collet17}, which are inherently limited to a maximum spatial resolution of about 250~nm and the scattering of corresponding wavelengths \cite{Sebastian15}, respectively, rendering them unable to access nanoscale waves and devices. Another commonly used experimental technique for studying spin waves is all electrical spin-wave spectroscopy using vector network analyzers \cite{Yu16, Bailleul03}. While this method is not limited by the wavelength, it does not allow for a direct imaging of spin waves. It also needs comparably large samples to achieve sufficient signal to noise ratio \cite{deLoubens07}, limiting its access to nanoscale devices. Thus, from both a fundamental and applications perspective there is a clear need for the spatially resolved detection of sub-250~nm spin waves.

Time-resolved scanning transmission x-ray microscopy (TR-STXM) is a technique that is able to meet these requirements \cite{VanWaey06, Nolle12, Noske14}. Magnetic phenomena can be routinely studied with spatial and stroboscopic temporal resolutions down to 20~nm and 50~ps respectively. Spin waves in metallic samples prepared as thin films on x-ray transparent silicon nitride (SiN) membranes have already been imaged successfully \cite{Dieterle17, Wintz16, Graefe17, Kammerer11, Gross19}. But the lack of x-ray transparency in the bulk substrates of single-crystalline systems like YIG films on gadolinium gallium garnet (GGG) requires an appropriate thinning route for STXM investigations \cite{Simmendinger18, Eisebitt17}. Therefore in the present work a thin sheet of YIG of the order of 185~nm thickness has been sliced out of a YIG thin film and its GGG substrate. The lamella was subsequently put onto an x-ray transparent SiN membrane (\textit{cf.} section \ref{Methods}). We present TR-STXM measurements in YIG, which provide a new view on the rich and complex scenario of the spin wave characteristics, their interactions and coexistence in the nm range of this pivotal model system for design and understanding of future magnonic/spintronic applications.

\section{Methods \label{Methods}}

A YIG film of 185 nm thickness was grown by liquid phase epitaxy (LPE) on (111)-oriented GGG \cite{Dubs17}. Ferromagnetic resonance measurements showed a saturation magnetization of $M_S=(143\pm2)\;\mathrm{kA/m}$ and a Gilbert damping coefficient of $\alpha=1.3\cdot 10^{-4}$, which both agree well with typical values for YIG films in literature \cite{Dubs17, Pirro14,Krysztofik17,Serga10}. The film was subsequently processed using a "FEI Dual Beam System Helios NanoLab 460F1" focused ion beam (FIB). A dedicated Ga$^{+}$ ion milling routine \cite{ErnstRuska16} resulted in a lamella of 13x5x0.185 $\mathrm{\mu m ^3}$ of YIG with less then 150 nm GGG attached to it. Afterwards, an "Omniprobe" micromanipulator was used to transfer the lamella to a standard SiN membrane, where it was centered on a copper microstrip antenna ($2\ \mathrm{\mu m}$ width and 200 nm thickness) and fixated with carbon. The copper microstrip was fabricated prior to the fixation of the lamella by a combination of electron beam lithography, thermal copper evaporation and lift-off processing.  

Measurements have been carried out at the MAXYMUS end station located at the UE46-PGM2 beam line at the BESSY II synchrotron radiation facility of Helmholtz-Zentrum Berlin. Circularly polarized x-rays were focused to 20~nm by a Fresnel zone plate. The X-ray magnetic circular dichroism (XMCD) effect \cite{Schuetz87} was used as magnetic contrast mechanism for imaging. For the x-ray energy the iron L$_3$-absorption edge was chosen, where the maximum magnetic signal fidelity was found at $(708\pm 0.3)\;\mathrm{eV}$ as a balance between XMCD strength and transmitted intensity \cite{Krichevtsov17}. The sample was mounted in normal incidence geometry, sensitive to the out-of-plane magnetization component. A quadrupole permanent magnet system provided an in-plane magnetic bias field in the range of $\pm 250\ \mathrm{mT}$ \cite{Nolle12}. Spin waves were excited by the magnetic field of an RF-current flowing through the copper stripline (\textit{cf.} figure \ref{fig1}).

 Time-resolution has been achieved by using a stroboscopic pump-and-probe technique that reaches a resolution of around 50~ps during the synchrotron's regular multibunch mode operation \cite{Noske14}. The raw movies from TR-STXM were normalized to enhance the dynamics. A pixel-wise fast Fourier transform (FFT) in the time-domain was subsequently used to obtain the local spin wave amplitude and phase \cite{Numpy, Matplotlib}, which were then used to visualize the waves in  HSV (hue, saturation, value) color space (\textit{c.f.} figure \ref{fig1}). A two-dimensional FFT in space was utilized to determine the corresponding wave vectors. See also the paper of Gro{\ss} et al. \cite{Gross19} for more details on this.

\section{Results \label{Results}}
\subsection{Experimental results \label{ExpRes}}

As a first step a continuous RF-current in the frequency range of $1.4$ to $3.0 ~\mathrm{GHz}$ was used for excitation. Figure \ref{fig1} shows the sample architecture and images of dynamics measured at different frequencies with an external magnetic field of $\mu_0H_{ext} = 25~\mathrm{mT}$ applied parallel to the stripline. The picture on the left in the upper row shows the raw x-ray intensity image of the lamella. Next to it are the frames from a time-resolved normalized movie at $1.6\ \mathrm{GHz}$ excitation frequency arranged in a time series. The frames show dynamic changes in the normal magnetization component as gray scale contrast. The vertical wavefronts of the BV type waves can clearly be seen, as well as their horizontal propagation between the time frames. From such movies, the visual representation shown in the images in the lower row have been obtained, showing the color coded Fourier amplitude and phase at each pixel (\textit{c.f.} section \ref{Methods}). As expected for spin waves in a thin film, a transition from wave front orientation normal to the external field (BV geometry) towards orientation parallel to the external field (DE geometry) can be observed when the frequency is raised \cite{Kalinikos86}. As is apparent in the first image ($1.4 ~ \mathrm{GHz}$) of the sequence, for the lowest frequencies the spin waves were confined to the sample edges due to the locally reduced effective field because of demagnetization effects as previously described in literature \cite{Jorzick02,Bayer03,Puszkar05}. The area of confinement extended from the edges to between 0.64~and 1.2~$\mathrm{\mu m}$ into the sample. In the second image (1.6~GHz), which corresponds to the time series above, two coexisting BV waves of different wavelengths ($\lambda_1=1.9 ~\mathrm{\mu m}$ and $\lambda_2=0.37 ~\mathrm{\mu m}$) are visible. Likewise, in the last two images ($2.5 \: \mathrm{and}\: 2.7 ~ \mathrm{GHz}$ respectively), where the waves are fully in DE orientation, two DE modes of different wavelengths appear, coexisting at the same frequency (at $f=2.7 ~\mathrm{GHz}$: $\lambda_1=2.8 ~\mathrm{\mu m}$ and $\lambda_2=0.53 ~\mathrm{\mu m}$). 

In a second step, excitation was changed from continuous sine wave to short bursts (\textit{cf.} figure \ref{fig2}, upper part). These bursts excited a broad spectrum of frequencies and, thus, a multitude of spin wave modes simultaneously (one normalized movie is attached as supplemental material). The center frequency of $f = 2~\mathrm{GHz}$ was chosen corresponding to the previously identified modes and thereby to cover a rich spin wave spectrum. The length of the burst was set to one sine period, or $\tau = 480~\mathrm{ps}$, while the downtime to the next burst was set to 31 sine periods, or approximately $\tau' = 15~\mathrm{ns}$. Single frequency components were isolated by a Fourier transform and, as previously, visualized in figure \ref{fig2}. The behavior is very similar to the continuous wave experiments shown in figure \ref{fig1}, especially for the direct comparison with the series in the middle row measured at the same field strength ($\mu_0H_{ext} = 25~\mathrm{mT}$). A transition from the BV to the DE propagation geometry at higher frequencies can be seen at all three magnetic bias field strengths. In agreement with theory \cite{Stancil09, Serga10}, the spin wave spectra, and hence the transition point, shift towards higher frequencies for increasing external fields. 

As the antenna was oriented for DE geometry excitation, its field is unlikely to be the primary source of the non-DE modes in the lamella. This point is reinforced by the observation, that those waves do not originate in the antenna's vicinity and rather from the lamella's edges, making reflections and the aforementioned edge demagnetization fields the probable causes. Especially for the BV geometry waves the observed transition from edge confined modes to sample-wide BV modes hints at the edge fields as source. The aforementioned secondary DE mode (short wavelength) also appears to originate from the upper and lower sample edge rather than from the antenna region.

\subsection{Analytical theory \label{TheoRes}}
To identify the specific spin waves that have been measured, their dispersion $f(k)$, where $f$ is the frequency and $k$ is the magnitude of the wavevector $\boldsymbol{k}$, was determined by a two-dimensional FFT (\textit{cf.} section \ref{Methods}). The focus was put on the two dominant wave orientations, namely the BV geometry ($\theta=0^{\circ}$) and the DE geometry ($\theta=90^{\circ}$), with $\theta$ being the angle between $\boldsymbol{k}$ and the equilibrium magnetization. Pairs of $f$ and $k$ were accordingly sorted by their $\theta$-values and compared to analytical models of basic spin wave modes in thin YIG films. A  model for spin waves in a thin ferromagnetic layer can be found in the work of Kalinikos and Slavin \cite{Kalinikos86}, taking the following approach.

An isotropic ferromagnetic film is considered, that is laterally infinite and of finite thickness $d$ along the $z$ axis ($z \in \left[-d/2, d/2\right]$). The film is magnetized in-plane by a magnetic bias field. A plane spin wave with a non-uniform vector amplitude $\boldsymbol{m} (z)$ is assumed to propagate in the film plane in the arbitrary $\zeta$ direction:

\begin{equation}
\boldsymbol{m}(z, \zeta, t)=\boldsymbol{m}(z) exp\left[i(\omega t-k\zeta)\right] \label{eqplanewave}
\end{equation}

\noindent where $t$ is time and $\omega=2\pi f$. The amplitude $\boldsymbol{m} (z)$ is then expanded into an infinite series of complete orthogonal vector functions. For this, the eigenfunctions of the second-order exchange differential operator satisfying the appropriate exchange boundary conditions, are chosen. For zero surface anisotropy (unpinned surface spins), which will be assumed from here on, this gives:

\begin{equation}
\boldsymbol{m}(z) \propto \sum_{n}{} \boldsymbol{m}_{n} cos \left[\kappa_n(z+\frac{d}{2})\right] \label{eqseries}
\end{equation}
 
\noindent where $\kappa_n = \frac{n \pi}{d}$, $n \in \mathbb{N}_0$, represents a standing wave component perpendicular to the film plane. Using equation \ref{eqseries}, the following infinite system of algebraic equations can be obtained from the well-known Landau-Lifshitz equation of motion:

\begin{equation}
-i \frac{\omega}{\omega_M}\boldsymbol{m}_n = \sum_{n'}{} \hat{W}_{n n'}\boldsymbol{m}_{n'} \label{eqmatrix}
\end{equation}

\noindent where $\omega_M =\gamma M_S$, $\gamma$ the gyromagnetic ratio and $M_S$ is the saturation magnetization. This corresponds to equation (22) in the source paper \cite{Kalinikos86}, where details on the square matrix $\hat{W}$ can also be found. The eigenvalues of this system give the frequency of the in-plane propagating spin wave modes of the film. The mode order $n$ here represents a standing spin wave component along the film thickness given by $\kappa_n$. For $k=0$ the mode coincides with the $n$-th order PSSW. This results in the amplitude profile $m(z)$ having $n$ nodes along the film thickness. If only the diagonal parts ($n=n'$) of $\hat{W}$ are considered, which means that interactions between modes of different orders are neglected, an approximate dispersion equation can be explicitly formulated \cite{Kalinikos86}:

\begin{multline}	
f_n=\frac{\gamma\mu_0}{2\pi}\biggr\{\left(H+\frac{2A}{\mu_0M_S}K^2\right)\times \\ \times\left(H+\frac{2A}{\mu_0M_S}K^2+M_SF_{nn}\right)\biggr\}^{1/2}\label{eqdisp}
\end{multline}		
  
with $K^2=k^2+\kappa_n^2$ and the element of the dipole-dipole matrix:

\begin{equation}
F_{nn}=1-P_{nn}cos^2\theta+P_{nn}(1-P_{nn})sin^2\theta \frac{M_S}{H+\frac{2A}{\mu_0M_S}K^2} \label{eqFnn}
\end{equation}

where $H$ is the magnitude of the magnetic field, $\mu_0$ the vacuum permeability, $A$ the exchange constant and $\theta$ the angle between the magnetization and $\boldsymbol{k}$. For the fundamental zero-order mode (uniform thickness profile, no PSSW-component) $P_{00}=1-\frac{1-e^{-kd}}{kd}$ (see appendix of the original paper\cite{Kalinikos86}). The zero-order equation gives the dispersions of the fundamental DE and BV modes at $\theta=90^{\circ}$ and $\theta=0^{\circ}$, respectively \footnote{Note that in the original Damon-Eshbach theory a non-uniform thickness profile is considered while the influence of exchange is neglected \cite{Damon65}.}.

\subsection{Comparing theory and experimental data \label{CompRes}}

\begin{figure*}[t]
	\includegraphics[width=0.95\textwidth]{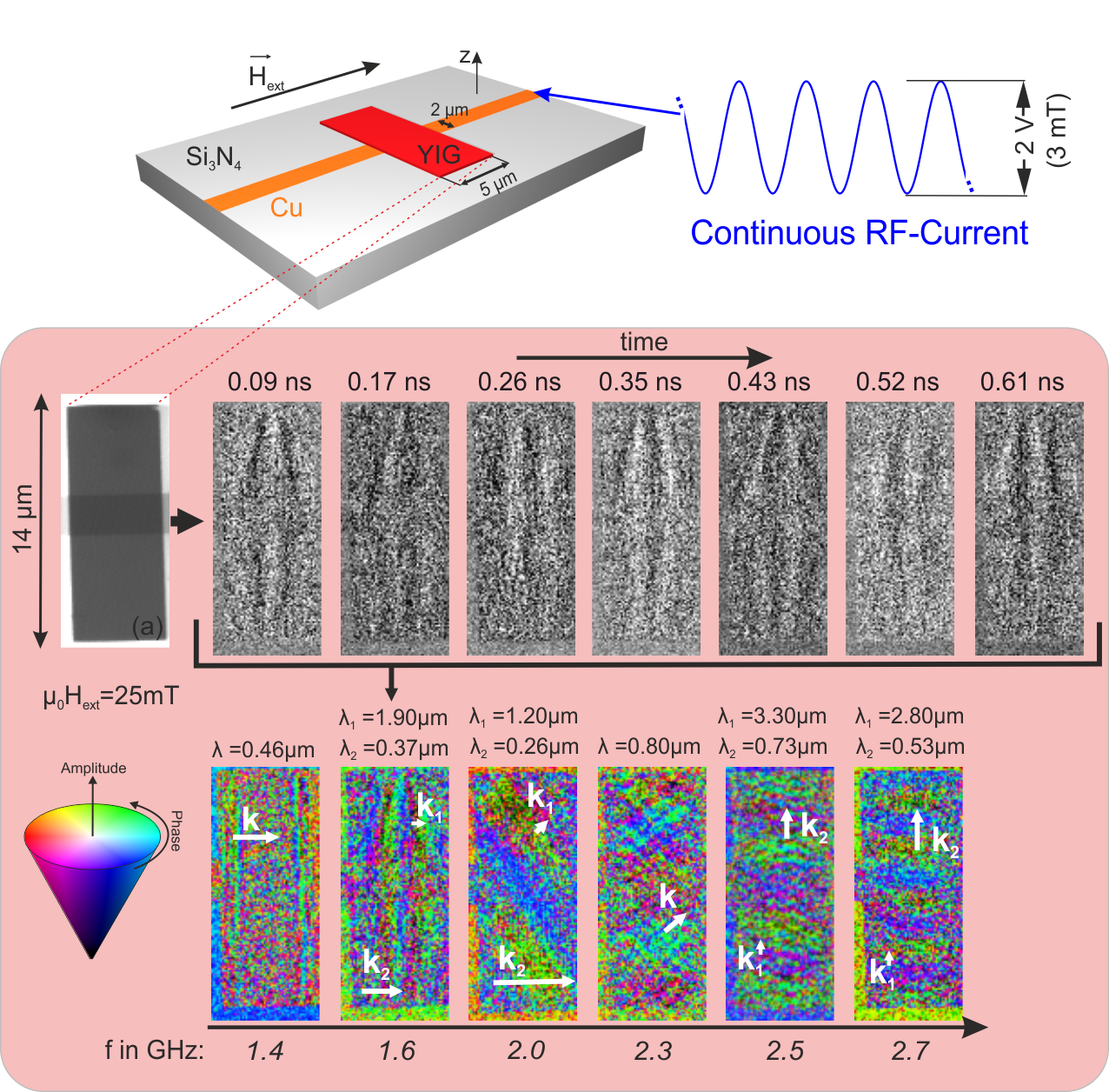}
	\caption{Upper part: Schematics of the sample used for the experiments. Gray: Silicon nitride membrane (Silson Ltd). Red cuboid: YIG Lamella, dimensions: $13 \times 5 \times 0.185 ~ \mathrm{\mu m^3}$. The magnetic bias field $\boldsymbol{H_{ext}}$ was oriented in the sample plane parallel to the copper stripline. Lower part: TR-STXM measurements of the sample at $\mu_0H_{ext} = 25~\mathrm{mT}$ and frequencies from $1.4$ to $2.7 ~\mathrm{GHz}$. Picture (a) in the upper row shows a raw x-ray image of the lamella (dark gray rectangle) and the stripline. The image series next to it displays the frames of a time-resolved movie at $1.6\ \mathrm{GHz}$, showing the normalized (see section \ref{Methods}) out-of-plane magnetization in arbitrary units. The color images in the lower row have been obtained from such movies by gaining the local Fourier amplitude and phase (\textit{c.f.} section \ref{Methods}) of each pixel's time-evolution and visualizing it in the HSV color space (color code on the upper left). Wave fronts visibly change from backward volume orientation, through diagonal intermediate states, to Damon-Eshbach direction as the frequency increases. Wave vector directions are indicated in the images with the corresponding wave lengths $\lambda$ stated above. \label{fig1}}.
\end{figure*}

\begin{figure*}[t]
	\includegraphics[width=\textwidth]{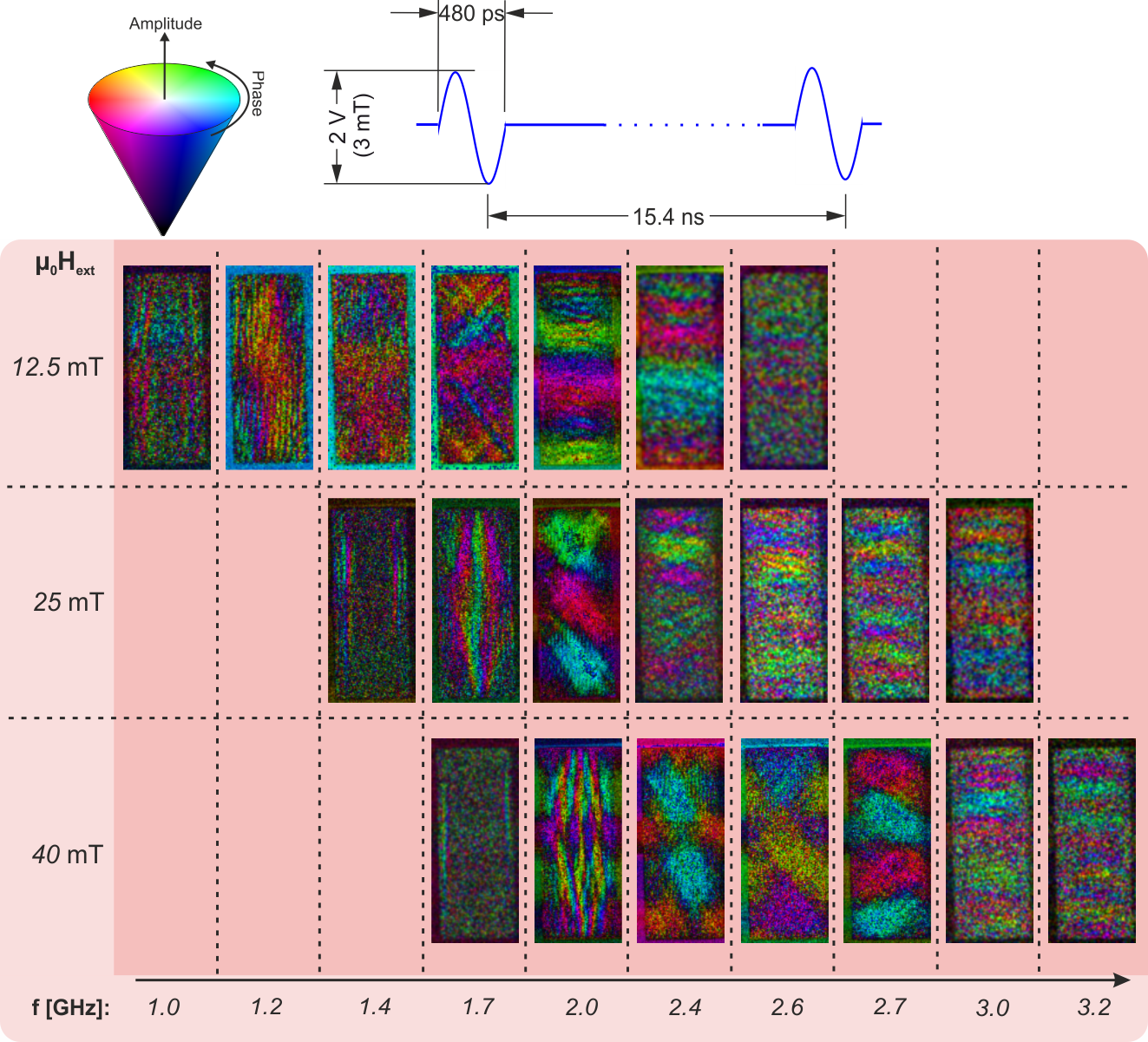}
	\caption{Top part: RF-burst signal used for excitation of the sample (Duration: 480 ps, repetition time: 15.4 ns, voltage amplitude: 2V). Main part: Results of burst measurements analogue to figure \ref{fig1} at three different magnetic bias fields $\boldsymbol{H_{ext}}$. Spin waves shift from the backward volume to the Damon-Eshbach propagation geometry as the frequency is raised, the transition point and general spectrum shifting to higher frequencies at greater field strength. \label{fig2}}
\end{figure*}

\begin{figure*}[t]
	\includegraphics[width=0.9\textwidth]{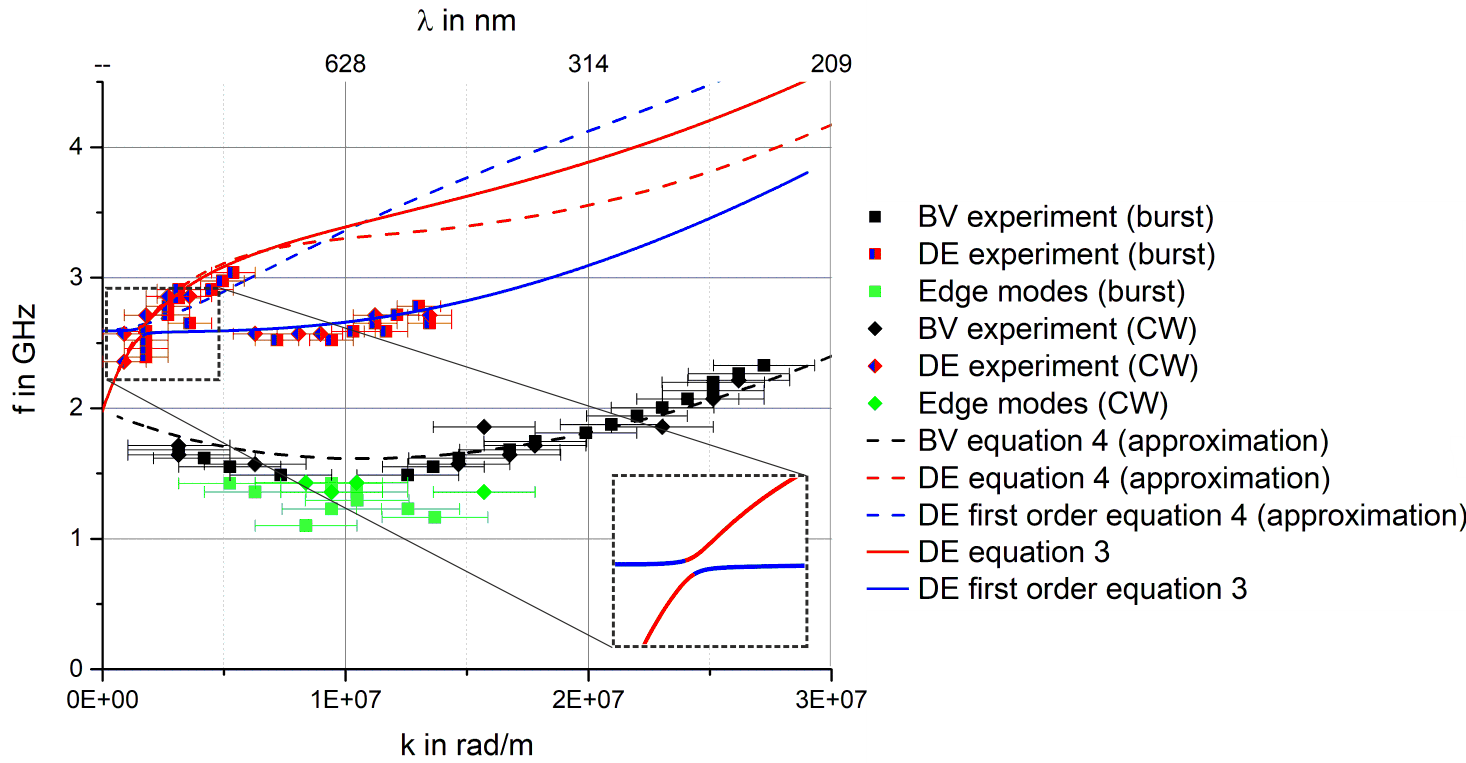}
	\caption{Plot of experimental dispersion data at $\mu_0H_{ext}=25~\mathrm{mT}$ for both continuous wave experiments (\textit{c.f.} figure \ref{fig1}) and the RF-burst experiment (\textit{c.f.} figure \ref{fig2}). Dots represent measured data sorted by propagation direction of the waves. Black and green dots show backward volume (BV) propagation ($\theta=0^\circ$) with the green dots marking those confined to the sample edges. The red-blue dots represent Damon-Eshbach(DE) propagation ($\theta=90^\circ$). Dashed lines show theoretical dispersion calculated using the approximate equation \ref{eqdisp} (no hybridization). The red and blue lines show the zero and first order DE dispersions, while the black line stands for the BV mode. The solid lines represent DE dispersion based on numerical calculations according to equation \ref{eqmatrix}, taking into account the hybridization of modes. The inset shows a magnification of the avoided crossing region of the two branches.\label{figdisp}}
\end{figure*}

In figure \ref{figdisp} the spin wave dispersion relation as deduced from the experimental data of both continuous wave and burst excitations for $\mu_0 H_{ext}=25\ \mathrm{mT}$ is shown. As figure \ref{fig1} and \ref{fig2} already suggested, it is not possible to distinguish between the dispersions measured for the two different excitation schemes, as the two data sets almost perfectly overlap. In a first step the corresponding theoretical dispersion curves based on the approximate equation (\ref{eqdisp}) were calculated ($M_S$ given in section \ref{Methods}, $A=0.36\cdot 10^{-11} ~\mathrm{J/m}$ \cite{Klingler15}) and plotted in figure \ref{figdisp} as dashed lines. The $\theta=0^\circ$-waves fit very well with the calculated fundamental BV dispersion curve (dashed black line). The influence of the exchange interaction becomes clear by means of the curve changing to a positive slope beyond $k\approx 1\cdot 10^7$~rad/m (compare to exchange free curves in Ref. \cite{Serga10, Stancil09}). This also explains the two coexisting BV waves mentioned in figure \ref{fig1}, as they originate from the branches of the curve left and right of the apex respectively. Thus, it appears that equation (\ref{eqdisp}) is a valid approximation for BV waves in this sample, at least in the wavelength range covered here. There seems to be no significant hybridization with higher order BV modes and or influence of the confined sample geometry besides them originating from the lateral edges. The edge modes, shown by the green dots, come close to the BV dispersion with a downward frequency shift of about 200~MHz. In order to explain this difference through edge demagnetization effects, a reduction of the effective field to about 15~$\mathrm{mT}$ would be necessary. According to micromagnetic simulations of the lamella's demagnetizing field this is a reasonable value at the edges.

The $\theta=90^\circ$-waves on the other hand appear to belong to two separate dispersion branches that coexist in the area between $f=2.5\ \mathrm{to}\ 2.9~\mathrm{GHz}$. As mentioned earlier (\textit{cf.} figure \ref{fig1}) two modes at $\theta=90^\circ$ can be seen simultaneously in the last images ($f\geq2.5\ \mathrm{GHz}$), which already hints at this behaviour. While the analytically approximated fundamental DE dispersion (dashed red line) fits the longer wavelength mode, the shorter wavelength one has to belong to a different spin wave mode featuring the same propagation direction. Obvious candidates for this are DE modes of higher orders. The dashed blue line in figure \ref{figdisp} represents the diagonal approximation for the first order thickness mode ($n=1$) by equation (\ref{eqdisp}). It is apparent that this approximation, \textit{i.e.} neglecting hybridization between different mode orders, is insufficient to describe the DE first order thickness mode in this particular system. 
  
Thus, in a second step, numerical calculations of the zero and first order DE dispersions have been carried out using the more accurate equation system (\ref{eqmatrix}) and considering the non-diagonal terms $n\neq n'$ of the matrix. The results are shown as solid lines (red and blue) in figure \ref{figdisp} and they notably diverge from the dashed analytical curves, while they agree very well with the experimental data. This strongly suggests that the two modes indeed hybridize. A closer look to the modes' crossing point (inset in figure \ref{figdisp}) supports this, as the presence of hybridization effects results in a band splitting. However, comparing the dashed and solid curves, it can be seen that the influence of the modes' mutual interaction reaches well beyond the crossing point. This agrees with observations made previously in 80 to 100~nm thick permalloy films \cite{Dieterle17}. Due to this, even the dispersion of the fundamental DE wave clearly diverges from the approximation of equation (\ref{eqdisp}) below wavelengths of 600~nm. This stresses the importance of using the extended calculation when going below $1\;\mu\mathrm{m}$ wavelength.

\subsection{Micromagnetic simulation \label{SimRes}}

\begin{figure}[h]
	\includegraphics[width=0.5\textwidth]{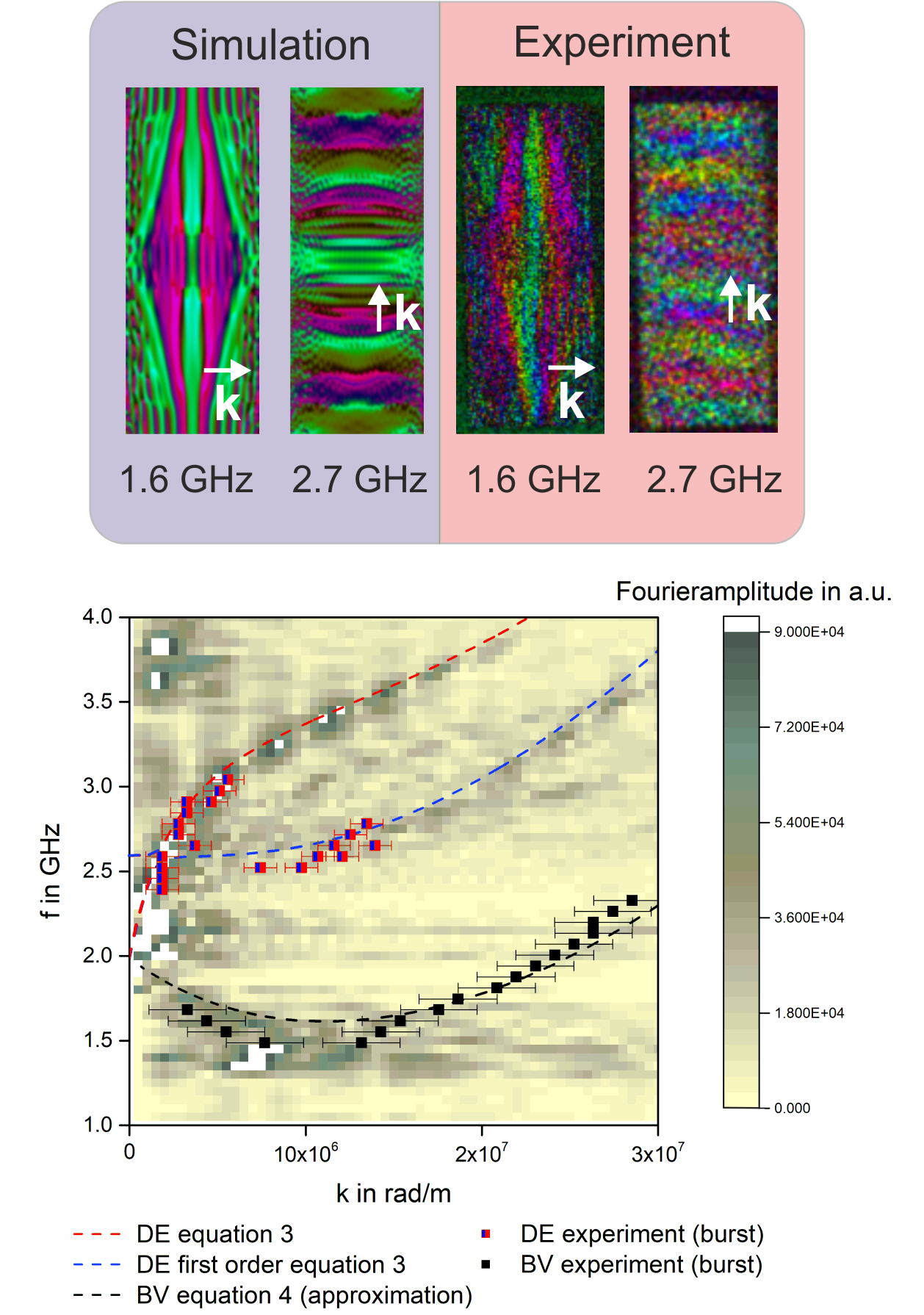}
	\caption{Upper part: Direct comparison of Fourier images of the experiment and the micromagnetic simulation at $\mu_0 H = 25~\mathrm{mT}$. Lower part: Heatmap of the spatial Fourier transform of the simulation for the two main directions $\theta=0^\circ$ and $\theta=90^\circ$ with the corresponding experimental data (dots) and theoretical dispersion relations (dashed lines) for the Damon-Eshbach modes of order zero and one based on equation \ref{eqmatrix}, as well as for the backward volume mode based on equation \ref{eqdisp}.\label{figsim1}}
\end{figure}

Finally, a micromagnetic simulation was carried out using the "MuMax 3" software developed by Vansteenkiste et al. \cite{Mumax}. For the simulation, the same material parameters as for the dispersion calculations were used, and external field of $\mu_0H=25\mathrm{mT}$ was considered together with an RF-burst excitation similar to the experimental one with a field amplitude of $3\ \mathrm{mT}$. Figure \ref{figsim1} shows direct comparisons of Fourier images from simulation and experiment at two different frequencies. Results for the dispersion along the main directions $\theta=0^\circ$ and $\theta=90^\circ$ are depicted in the lower part of figure \ref{figsim1} and show reasonable agreement with the physical measurements (dots) and theory accounting for hybridization (dashed lines). This gives reason to assume that the simulation is a good representation of the experiment and that the simulation results beyond the experimentally covered region represent a viable extrapolation. The agreement of simulation and theory in such advanced regions further supports the theoretical approach taken.  

The simulation also highlights another important point. As figure \ref{figsim1} shows, the dispersion curves in theory and simulation continue beyond the experimentally observed data range. This is especially apparent for the DE-waves, which stay well below the wavenumbers measured in the BV-geometry, that themselves reach a limit at $k\approx 3 \cdot 10^{-7}$~rad/m ($\lambda \approx 200$~nm). Physically, for every antenna-like spin wave source there is a sharply diminishing efficiency of excitation for wavelengths below the source's width. This limits the wavevectors that can be excited by the source at a given energy input. Since the BV-waves are likely excited by the approximately $1 \mathrm{\mu m}$ wide demagnetization fields on the lateral edges, the limit for them is lower than for the zero order DE-waves, which are excited by the $2 \mathrm{\mu m}$ wide copper antenna. Hence the occurence of much shorter waves in BV-geometry. 

\section{Conclusions \label{Conclusions}}

In summary, spin waves of wavelengths down to 200~nm have been directly imaged in YIG using TR-STXM. For this, a nearly freestanding lamella was fabricated from a YIG film by focused ion beam preparation. Spin wave modes of various directions in the sample plane have been recorded as a function of frequency and external magnetic field. TR-STXM enabled the simultaneous determination of their spatial properties, like wavefront shape, propagation direction or confinement to certain regions (\textit{e.g.} the edge), and of the waves time domain features. Dynamics were excited by continuous single frequency RF-fields as well as by broad band RF-bursts. The observed BV waves agree very well with a simple diagonal approximation of the analytical expression for the dispersion relation \cite{Kalinikos86}. This approach still held reasonably for the zero order DE mode up to $k\approx6\cdot10^{-6}$~rad/m. However, a second DE dispersion branch was observed, leading to the coexistence of two DE modes with strongly different wavelengths in the frequency range between 2.5 and 2.9~GHz. The diagonal approximation does not correctly describe the second mode, neither as DE zero order nor as its first higher order thickness mode. A more rigorous numerical calculation based on the full set of equations \cite{Kalinikos86} was necessary and provided an excellent match with the experimental findings. It can be concluded that hybridization between different mode orders plays a major role in this system for the formation of spin waves propagating in the DE geometry. Micromagnetic simulations have been done and fit well with the experimental data and the calculated dispersions, indicating their potential to predict the observed magnonic scenario in the system studied. While all analytic calculations assumed a laterally infinite film, it appears that the measured wavelengths were sufficiently small compared to the dimensions of the sample to still warrant this assumption. 

The presented work demonstrates that TR-STXM is a powerful and versatile tool for high resolution imaging of magnetization dynamics in real space and time domain. It clearly demonstrates its applicability to YIG thin films, making these accessible to space and time resolved spin wave studies beyond optical resolution limits. This opens up a pathway to directly image nanoscaled spin dynamics in YIG and other single crystalline materials and will have an important impact for fundamental magnonic research and applications in nano devices.

\begin{acknowledgments}
	We thank HZB for the allocation of synchrotron radiation beamtime. Michael Bechtel is gratefully acknowledged for support during beamtimes. J.B. is supported from the European Union’s Horizon 2020 research and innovation programme under the Marie Skłodowska-Curie grant agreement No.66566. C.D. acknowledges the financial support by the Deutsche Forschungsgemeinschaft (DU 1427/2-1). A.N.S. was supported by the Grant Nos. EFMA-1641989 and ECCS-1708982 from the National Science Foundation (NSF) of the USA, and by the Defense Advanced Research Projects Agency (DARPA) M3IC Grant under Contract No. W911-17-C-0031.
 
\end{acknowledgments}

\clearpage

\clearpage
\bibliography{Bibliography}

\begin{thebibliography}{49}%
\makeatletter
\providecommand \@ifxundefined [1]{%
 \@ifx{#1\undefined}
}%
\providecommand \@ifnum [1]{%
 \ifnum #1\expandafter \@firstoftwo
 \else \expandafter \@secondoftwo
 \fi
}%
\providecommand \@ifx [1]{%
 \ifx #1\expandafter \@firstoftwo
 \else \expandafter \@secondoftwo
 \fi
}%
\providecommand \natexlab [1]{#1}%
\providecommand \enquote  [1]{``#1''}%
\providecommand \bibnamefont  [1]{#1}%
\providecommand \bibfnamefont [1]{#1}%
\providecommand \citenamefont [1]{#1}%
\providecommand \href@noop [0]{\@secondoftwo}%
\providecommand \href [0]{\begingroup \@sanitize@url \@href}%
\providecommand \@href[1]{\@@startlink{#1}\@@href}%
\providecommand \@@href[1]{\endgroup#1\@@endlink}%
\providecommand \@sanitize@url [0]{\catcode `\\12\catcode `\$12\catcode
  `\&12\catcode `\#12\catcode `\^12\catcode `\_12\catcode `\%12\relax}%
\providecommand \@@startlink[1]{}%
\providecommand \@@endlink[0]{}%
\providecommand \url  [0]{\begingroup\@sanitize@url \@url }%
\providecommand \@url [1]{\endgroup\@href {#1}{\urlprefix }}%
\providecommand \urlprefix  [0]{URL }%
\providecommand \Eprint [0]{\href }%
\providecommand \doibase [0]{http://dx.doi.org/}%
\providecommand \selectlanguage [0]{\@gobble}%
\providecommand \bibinfo  [0]{\@secondoftwo}%
\providecommand \bibfield  [0]{\@secondoftwo}%
\providecommand \translation [1]{[#1]}%
\providecommand \BibitemOpen [0]{}%
\providecommand \bibitemStop [0]{}%
\providecommand \bibitemNoStop [0]{.\EOS\space}%
\providecommand \EOS [0]{\spacefactor3000\relax}%
\providecommand \BibitemShut  [1]{\csname bibitem#1\endcsname}%
\let\auto@bib@innerbib\@empty
\bibitem [{\citenamefont {Demokritov}\ \emph {et~al.}(2006)\citenamefont
  {Demokritov}, \citenamefont {Demidov}, \citenamefont {Dzyapko}, \citenamefont
  {Melkov}, \citenamefont {Serga}, \citenamefont {Hillebrands},\ and\
  \citenamefont {Slavin}}]{Demokritov06}%
  \BibitemOpen
  \bibfield  {author} {\bibinfo {author} {\bibfnamefont {S.~O.}\ \bibnamefont
  {Demokritov}}, \bibinfo {author} {\bibfnamefont {V.~E.}\ \bibnamefont
  {Demidov}}, \bibinfo {author} {\bibfnamefont {O.}~\bibnamefont {Dzyapko}},
  \bibinfo {author} {\bibfnamefont {G.~A.}\ \bibnamefont {Melkov}}, \bibinfo
  {author} {\bibfnamefont {A.~A.}\ \bibnamefont {Serga}}, \bibinfo {author}
  {\bibfnamefont {B.}~\bibnamefont {Hillebrands}}, \ and\ \bibinfo {author}
  {\bibfnamefont {A.~N.}\ \bibnamefont {Slavin}},\ }\href {\doibase
  10.1038/nature05117} {\bibfield  {journal} {\bibinfo  {journal} {Nature}\
  }\textbf {\bibinfo {volume} {443}},\ \bibinfo {pages} {430} (\bibinfo {year}
  {2006})}\BibitemShut {NoStop}%
\bibitem [{\citenamefont {Bozhko}\ \emph {et~al.}(2016)\citenamefont {Bozhko},
  \citenamefont {Serga}, \citenamefont {Clausen}, \citenamefont {Vasyuchka},
  \citenamefont {Heussner}, \citenamefont {Melkov}, \citenamefont {Pomyalov},
  \citenamefont {L’vov},\ and\ \citenamefont {Hillebrands}}]{Bozhko16}%
  \BibitemOpen
  \bibfield  {author} {\bibinfo {author} {\bibfnamefont {D.~A.}\ \bibnamefont
  {Bozhko}}, \bibinfo {author} {\bibfnamefont {A.~A.}\ \bibnamefont {Serga}},
  \bibinfo {author} {\bibfnamefont {P.}~\bibnamefont {Clausen}}, \bibinfo
  {author} {\bibfnamefont {V.~I.}\ \bibnamefont {Vasyuchka}}, \bibinfo {author}
  {\bibfnamefont {F.}~\bibnamefont {Heussner}}, \bibinfo {author}
  {\bibfnamefont {G.~A.}\ \bibnamefont {Melkov}}, \bibinfo {author}
  {\bibfnamefont {A.}~\bibnamefont {Pomyalov}}, \bibinfo {author}
  {\bibfnamefont {V.~S.}\ \bibnamefont {L’vov}}, \ and\ \bibinfo {author}
  {\bibfnamefont {B.}~\bibnamefont {Hillebrands}},\ }\href {\doibase
  10.1038/nphys3838} {\bibfield  {journal} {\bibinfo  {journal} {Nature
  Physics}\ }\textbf {\bibinfo {volume} {12}},\ \bibinfo {pages} {1057}
  (\bibinfo {year} {2016})}\BibitemShut {NoStop}%
\bibitem [{\citenamefont {Roldán-Molina}\ \emph {et~al.}(2017)\citenamefont
  {Roldán-Molina}, \citenamefont {Nunez},\ and\ \citenamefont
  {Duine}}]{Roldan_Molina17}%
  \BibitemOpen
  \bibfield  {author} {\bibinfo {author} {\bibfnamefont {A.}~\bibnamefont
  {Roldán-Molina}}, \bibinfo {author} {\bibfnamefont {A.~S.}\ \bibnamefont
  {Nunez}}, \ and\ \bibinfo {author} {\bibfnamefont {R.~A.}\ \bibnamefont
  {Duine}},\ }\href {\doibase 10.1103/PhysRevLett.118.061301} {\bibfield
  {journal} {\bibinfo  {journal} {Physical Review Letters}\ }\textbf {\bibinfo
  {volume} {118}},\ \bibinfo {pages} {061301} (\bibinfo {year}
  {2017})}\BibitemShut {NoStop}%
\bibitem [{\citenamefont {Serga}\ \emph {et~al.}(2010)\citenamefont {Serga},
  \citenamefont {Chumak},\ and\ \citenamefont {Hillebrands}}]{Serga10}%
  \BibitemOpen
  \bibfield  {author} {\bibinfo {author} {\bibfnamefont {A.~A.}\ \bibnamefont
  {Serga}}, \bibinfo {author} {\bibfnamefont {A.~V.}\ \bibnamefont {Chumak}}, \
  and\ \bibinfo {author} {\bibfnamefont {B.}~\bibnamefont {Hillebrands}},\
  }\href {http://stacks.iop.org/0022-3727/43/i=26/a=264002} {\bibfield
  {journal} {\bibinfo  {journal} {Journal of Physics D: Applied Physics}\
  }\textbf {\bibinfo {volume} {43}},\ \bibinfo {pages} {264002} (\bibinfo
  {year} {2010})}\BibitemShut {NoStop}%
\bibitem [{\citenamefont {Chumak}\ \emph {et~al.}(2015)\citenamefont {Chumak},
  \citenamefont {Vasyuchka}, \citenamefont {Serga},\ and\ \citenamefont
  {Hillebrands}}]{Chumak15}%
  \BibitemOpen
  \bibfield  {author} {\bibinfo {author} {\bibfnamefont {A.~V.}\ \bibnamefont
  {Chumak}}, \bibinfo {author} {\bibfnamefont {V.~I.}\ \bibnamefont
  {Vasyuchka}}, \bibinfo {author} {\bibfnamefont {A.~A.}\ \bibnamefont
  {Serga}}, \ and\ \bibinfo {author} {\bibfnamefont {B.}~\bibnamefont
  {Hillebrands}},\ }\href {\doibase 10.1038/nphys3347} {\bibfield  {journal}
  {\bibinfo  {journal} {Nature Physics}\ }\textbf {\bibinfo {volume} {11}},\
  \bibinfo {pages} {453} (\bibinfo {year} {2015})}\BibitemShut {NoStop}%
\bibitem [{\citenamefont {Stancil;}\ and\ \citenamefont
  {Prabhakar}(2009)}]{Stancil09}%
  \BibitemOpen
  \bibfield  {author} {\bibinfo {author} {\bibfnamefont {D.~D.}\ \bibnamefont
  {Stancil;}}\ and\ \bibinfo {author} {\bibfnamefont {A.}~\bibnamefont
  {Prabhakar}},\ }\href {\doibase 10.1007/978-0-387-77865-5} {\emph {\bibinfo
  {title} {Spin Waves, Theory and Applications}}},\ \bibinfo {edition} {1st}\
  ed.\ (\bibinfo  {publisher} {Springer US},\ \bibinfo {year} {2009})\ pp.\
  \bibinfo {pages} {XII, 348}\BibitemShut {NoStop}%
\bibitem [{\citenamefont {Damon}\ and\ \citenamefont {Vaart}(1965)}]{Damon65}%
  \BibitemOpen
  \bibfield  {author} {\bibinfo {author} {\bibfnamefont {R.~W.}\ \bibnamefont
  {Damon}}\ and\ \bibinfo {author} {\bibfnamefont {H.~v.~d.}\ \bibnamefont
  {Vaart}},\ }\href {\doibase 10.1109/PROC.1965.3747} {\bibfield  {journal}
  {\bibinfo  {journal} {Proceedings of the IEEE}\ }\textbf {\bibinfo {volume}
  {53}},\ \bibinfo {pages} {348} (\bibinfo {year} {1965})}\BibitemShut
  {NoStop}%
\bibitem [{\citenamefont {Liu}\ \emph {et~al.}(2015)\citenamefont {Liu},
  \citenamefont {Riley},\ and\ \citenamefont {Buchanan}}]{Liu15}%
  \BibitemOpen
  \bibfield  {author} {\bibinfo {author} {\bibfnamefont {H.~J.~J.}\
  \bibnamefont {Liu}}, \bibinfo {author} {\bibfnamefont {G.~A.}\ \bibnamefont
  {Riley}}, \ and\ \bibinfo {author} {\bibfnamefont {K.~S.}\ \bibnamefont
  {Buchanan}},\ }\href {\doibase 10.1109/LMAG.2015.2495162} {\bibfield
  {journal} {\bibinfo  {journal} {IEEE Magnetics Letters}\ }\textbf {\bibinfo
  {volume} {6}},\ \bibinfo {pages} {1} (\bibinfo {year} {2015})}\BibitemShut
  {NoStop}%
\bibitem [{\citenamefont {Klingler}\ \emph
  {et~al.}(2015{\natexlab{a}})\citenamefont {Klingler}, \citenamefont {Pirro},
  \citenamefont {Br\"acher}, \citenamefont {Leven}, \citenamefont
  {Hillebrands},\ and\ \citenamefont {Chumak}}]{Klingler15_2}%
  \BibitemOpen
  \bibfield  {author} {\bibinfo {author} {\bibfnamefont {S.}~\bibnamefont
  {Klingler}}, \bibinfo {author} {\bibfnamefont {P.}~\bibnamefont {Pirro}},
  \bibinfo {author} {\bibfnamefont {T.}~\bibnamefont {Br\"acher}}, \bibinfo
  {author} {\bibfnamefont {B.}~\bibnamefont {Leven}}, \bibinfo {author}
  {\bibfnamefont {B.}~\bibnamefont {Hillebrands}}, \ and\ \bibinfo {author}
  {\bibfnamefont {A.~V.}\ \bibnamefont {Chumak}},\ }\href {\doibase
  10.1063/1.4921850} {\bibfield  {journal} {\bibinfo  {journal} {Applied
  Physics Letters}\ }\textbf {\bibinfo {volume} {106}},\ \bibinfo {pages}
  {212406} (\bibinfo {year} {2015}{\natexlab{a}})}\BibitemShut {NoStop}%
\bibitem [{\citenamefont {Kalinikos}\ and\ \citenamefont
  {Slavin}(1986)}]{Kalinikos86}%
  \BibitemOpen
  \bibfield  {author} {\bibinfo {author} {\bibfnamefont {B.~A.}\ \bibnamefont
  {Kalinikos}}\ and\ \bibinfo {author} {\bibfnamefont {A.~N.}\ \bibnamefont
  {Slavin}},\ }\href {http://stacks.iop.org/0022-3719/19/i=35/a=014} {\bibfield
   {journal} {\bibinfo  {journal} {Journal of Physics C: Solid State Physics}\
  }\textbf {\bibinfo {volume} {19}},\ \bibinfo {pages} {7013} (\bibinfo {year}
  {1986})}\BibitemShut {NoStop}%
\bibitem [{\citenamefont {Bozhko}\ \emph {et~al.}(2017)\citenamefont {Bozhko},
  \citenamefont {Clausen}, \citenamefont {Melkov}, \citenamefont {L’vov},
  \citenamefont {Pomyalov}, \citenamefont {Vasyuchka}, \citenamefont {Chumak},
  \citenamefont {Hillebrands},\ and\ \citenamefont {Serga}}]{Bozhko17}%
  \BibitemOpen
  \bibfield  {author} {\bibinfo {author} {\bibfnamefont {D.~A.}\ \bibnamefont
  {Bozhko}}, \bibinfo {author} {\bibfnamefont {P.}~\bibnamefont {Clausen}},
  \bibinfo {author} {\bibfnamefont {G.~A.}\ \bibnamefont {Melkov}}, \bibinfo
  {author} {\bibfnamefont {V.~S.}\ \bibnamefont {L’vov}}, \bibinfo {author}
  {\bibfnamefont {A.}~\bibnamefont {Pomyalov}}, \bibinfo {author}
  {\bibfnamefont {V.~I.}\ \bibnamefont {Vasyuchka}}, \bibinfo {author}
  {\bibfnamefont {A.~V.}\ \bibnamefont {Chumak}}, \bibinfo {author}
  {\bibfnamefont {B.}~\bibnamefont {Hillebrands}}, \ and\ \bibinfo {author}
  {\bibfnamefont {A.~A.}\ \bibnamefont {Serga}},\ }\href {\doibase
  10.1103/PhysRevLett.118.237201} {\bibfield  {journal} {\bibinfo  {journal}
  {Physical Review Letters}\ }\textbf {\bibinfo {volume} {118}},\ \bibinfo
  {pages} {237201} (\bibinfo {year} {2017})}\BibitemShut {NoStop}%
\bibitem [{\citenamefont {Kreil}\ \emph {et~al.}(2018)\citenamefont {Kreil},
  \citenamefont {Bozhko}, \citenamefont {Musiienko-Shmarova}, \citenamefont
  {Vasyuchka}, \citenamefont {L’vov}, \citenamefont {Pomyalov}, \citenamefont
  {Hillebrands},\ and\ \citenamefont {Serga}}]{Kreil18}%
  \BibitemOpen
  \bibfield  {author} {\bibinfo {author} {\bibfnamefont {A.~J.~E.}\
  \bibnamefont {Kreil}}, \bibinfo {author} {\bibfnamefont {D.~A.}\ \bibnamefont
  {Bozhko}}, \bibinfo {author} {\bibfnamefont {H.~Y.}\ \bibnamefont
  {Musiienko-Shmarova}}, \bibinfo {author} {\bibfnamefont {V.~I.}\ \bibnamefont
  {Vasyuchka}}, \bibinfo {author} {\bibfnamefont {V.~S.}\ \bibnamefont
  {L’vov}}, \bibinfo {author} {\bibfnamefont {A.}~\bibnamefont {Pomyalov}},
  \bibinfo {author} {\bibfnamefont {B.}~\bibnamefont {Hillebrands}}, \ and\
  \bibinfo {author} {\bibfnamefont {A.~A.}\ \bibnamefont {Serga}},\ }\href
  {\doibase 10.1103/PhysRevLett.121.077203} {\bibfield  {journal} {\bibinfo
  {journal} {Physical Review Letters}\ }\textbf {\bibinfo {volume} {121}},\
  \bibinfo {pages} {077203} (\bibinfo {year} {2018})}\BibitemShut {NoStop}%
\bibitem [{\citenamefont {Serga}\ \emph {et~al.}(2014)\citenamefont {Serga},
  \citenamefont {Tiberkevich}, \citenamefont {Sandweg}, \citenamefont
  {Vasyuchka}, \citenamefont {Bozhko}, \citenamefont {Chumak}, \citenamefont
  {Neumann}, \citenamefont {Obry}, \citenamefont {Melkov}, \citenamefont
  {Slavin},\ and\ \citenamefont {Hillebrands}}]{Serga14}%
  \BibitemOpen
  \bibfield  {author} {\bibinfo {author} {\bibfnamefont {A.~A.}\ \bibnamefont
  {Serga}}, \bibinfo {author} {\bibfnamefont {V.~S.}\ \bibnamefont
  {Tiberkevich}}, \bibinfo {author} {\bibfnamefont {C.~W.}\ \bibnamefont
  {Sandweg}}, \bibinfo {author} {\bibfnamefont {V.~I.}\ \bibnamefont
  {Vasyuchka}}, \bibinfo {author} {\bibfnamefont {D.~A.}\ \bibnamefont
  {Bozhko}}, \bibinfo {author} {\bibfnamefont {A.~V.}\ \bibnamefont {Chumak}},
  \bibinfo {author} {\bibfnamefont {T.}~\bibnamefont {Neumann}}, \bibinfo
  {author} {\bibfnamefont {B.}~\bibnamefont {Obry}}, \bibinfo {author}
  {\bibfnamefont {G.~A.}\ \bibnamefont {Melkov}}, \bibinfo {author}
  {\bibfnamefont {A.~N.}\ \bibnamefont {Slavin}}, \ and\ \bibinfo {author}
  {\bibfnamefont {B.}~\bibnamefont {Hillebrands}},\ }\href {\doibase
  10.1038/ncomms4452} {\bibfield  {journal} {\bibinfo  {journal} {Nature
  Communications}\ }\textbf {\bibinfo {volume} {5}},\ \bibinfo {pages} {3452}
  (\bibinfo {year} {2014})}\BibitemShut {NoStop}%
\bibitem [{\citenamefont {Kaack}\ \emph {et~al.}(1999)\citenamefont {Kaack},
  \citenamefont {Jun}, \citenamefont {Nikitov},\ and\ \citenamefont
  {Pelzl}}]{Kaak99}%
  \BibitemOpen
  \bibfield  {author} {\bibinfo {author} {\bibfnamefont {M.}~\bibnamefont
  {Kaack}}, \bibinfo {author} {\bibfnamefont {S.}~\bibnamefont {Jun}}, \bibinfo
  {author} {\bibfnamefont {S.~A.}\ \bibnamefont {Nikitov}}, \ and\ \bibinfo
  {author} {\bibfnamefont {J.}~\bibnamefont {Pelzl}},\ }\href {\doibase
  https://doi.org/10.1016/S0304-8853(99)00339-X} {\bibfield  {journal}
  {\bibinfo  {journal} {Journal of Magnetism and Magnetic Materials}\ }\textbf
  {\bibinfo {volume} {204}},\ \bibinfo {pages} {90} (\bibinfo {year}
  {1999})}\BibitemShut {NoStop}%
\bibitem [{\citenamefont {Baek}\ \emph {et~al.}(2004)\citenamefont {Baek},
  \citenamefont {Ha}, \citenamefont {Lim},\ and\ \citenamefont {Lee}}]{Baek04}%
  \BibitemOpen
  \bibfield  {author} {\bibinfo {author} {\bibfnamefont {J.~S.}\ \bibnamefont
  {Baek}}, \bibinfo {author} {\bibfnamefont {S.~Y.}\ \bibnamefont {Ha}},
  \bibinfo {author} {\bibfnamefont {W.~Y.}\ \bibnamefont {Lim}}, \ and\
  \bibinfo {author} {\bibfnamefont {S.~H.}\ \bibnamefont {Lee}},\ }\href
  {\doibase doi:10.1002/pssa.200304554} {\bibfield  {journal} {\bibinfo
  {journal} {physica status solidi (a)}\ }\textbf {\bibinfo {volume} {201}},\
  \bibinfo {pages} {1806} (\bibinfo {year} {2004})}\BibitemShut {NoStop}%
\bibitem [{\citenamefont {Fischer}\ \emph {et~al.}(2017)\citenamefont
  {Fischer}, \citenamefont {Kewenig}, \citenamefont {Bozhko}, \citenamefont
  {Serga}, \citenamefont {Syvorotka}, \citenamefont {Ciubotaru}, \citenamefont
  {Adelmann}, \citenamefont {Hillebrands},\ and\ \citenamefont
  {Chumak}}]{Fischer17}%
  \BibitemOpen
  \bibfield  {author} {\bibinfo {author} {\bibfnamefont {T.}~\bibnamefont
  {Fischer}}, \bibinfo {author} {\bibfnamefont {M.}~\bibnamefont {Kewenig}},
  \bibinfo {author} {\bibfnamefont {D.~A.}\ \bibnamefont {Bozhko}}, \bibinfo
  {author} {\bibfnamefont {A.~A.}\ \bibnamefont {Serga}}, \bibinfo {author}
  {\bibfnamefont {I.~I.}\ \bibnamefont {Syvorotka}}, \bibinfo {author}
  {\bibfnamefont {F.}~\bibnamefont {Ciubotaru}}, \bibinfo {author}
  {\bibfnamefont {C.}~\bibnamefont {Adelmann}}, \bibinfo {author}
  {\bibfnamefont {B.}~\bibnamefont {Hillebrands}}, \ and\ \bibinfo {author}
  {\bibfnamefont {A.~V.}\ \bibnamefont {Chumak}},\ }\href {\doibase
  10.1063/1.4979840} {\bibfield  {journal} {\bibinfo  {journal} {Applied
  Physics Letters}\ }\textbf {\bibinfo {volume} {110}},\ \bibinfo {pages}
  {152401} (\bibinfo {year} {2017})}\BibitemShut {NoStop}%
\bibitem [{\citenamefont {Nembach}\ \emph {et~al.}(2013)\citenamefont
  {Nembach}, \citenamefont {Shaw}, \citenamefont {Boone},\ and\ \citenamefont
  {Silva}}]{Nembach13}%
  \BibitemOpen
  \bibfield  {author} {\bibinfo {author} {\bibfnamefont {H.~T.}\ \bibnamefont
  {Nembach}}, \bibinfo {author} {\bibfnamefont {J.~M.}\ \bibnamefont {Shaw}},
  \bibinfo {author} {\bibfnamefont {C.~T.}\ \bibnamefont {Boone}}, \ and\
  \bibinfo {author} {\bibfnamefont {T.~J.}\ \bibnamefont {Silva}},\ }\href
  {\doibase 10.1103/PhysRevLett.110.117201} {\bibfield  {journal} {\bibinfo
  {journal} {Physical Review Letters}\ }\textbf {\bibinfo {volume} {110}},\
  \bibinfo {pages} {117201} (\bibinfo {year} {2013})}\BibitemShut {NoStop}%
\bibitem [{\citenamefont {Adur}\ \emph {et~al.}(2014)\citenamefont {Adur},
  \citenamefont {Du}, \citenamefont {Wang}, \citenamefont {Manuilov},
  \citenamefont {Bhallamudi}, \citenamefont {Zhang}, \citenamefont {Pelekhov},
  \citenamefont {Yang},\ and\ \citenamefont {Hammel}}]{Adur14}%
  \BibitemOpen
  \bibfield  {author} {\bibinfo {author} {\bibfnamefont {R.}~\bibnamefont
  {Adur}}, \bibinfo {author} {\bibfnamefont {C.}~\bibnamefont {Du}}, \bibinfo
  {author} {\bibfnamefont {H.}~\bibnamefont {Wang}}, \bibinfo {author}
  {\bibfnamefont {S.~A.}\ \bibnamefont {Manuilov}}, \bibinfo {author}
  {\bibfnamefont {V.~P.}\ \bibnamefont {Bhallamudi}}, \bibinfo {author}
  {\bibfnamefont {C.}~\bibnamefont {Zhang}}, \bibinfo {author} {\bibfnamefont
  {D.~V.}\ \bibnamefont {Pelekhov}}, \bibinfo {author} {\bibfnamefont
  {F.}~\bibnamefont {Yang}}, \ and\ \bibinfo {author} {\bibfnamefont {P.~C.}\
  \bibnamefont {Hammel}},\ }\href
  {https://link.aps.org/doi/10.1103/PhysRevLett.113.176601} {\bibfield
  {journal} {\bibinfo  {journal} {Physical Review Letters}\ }\textbf {\bibinfo
  {volume} {113}},\ \bibinfo {pages} {176601} (\bibinfo {year}
  {2014})}\BibitemShut {NoStop}%
\bibitem [{\citenamefont {Talalaevskij}\ \emph {et~al.}(2017)\citenamefont
  {Talalaevskij}, \citenamefont {Decker}, \citenamefont {Stigloher},
  \citenamefont {Mitra}, \citenamefont {K\"orner}, \citenamefont {Cespedes},
  \citenamefont {Back},\ and\ \citenamefont {Hickey}}]{Talalaevskij17}%
  \BibitemOpen
  \bibfield  {author} {\bibinfo {author} {\bibfnamefont {A.}~\bibnamefont
  {Talalaevskij}}, \bibinfo {author} {\bibfnamefont {M.}~\bibnamefont
  {Decker}}, \bibinfo {author} {\bibfnamefont {J.}~\bibnamefont {Stigloher}},
  \bibinfo {author} {\bibfnamefont {A.}~\bibnamefont {Mitra}}, \bibinfo
  {author} {\bibfnamefont {H.~S.}\ \bibnamefont {K\"orner}}, \bibinfo {author}
  {\bibfnamefont {O.}~\bibnamefont {Cespedes}}, \bibinfo {author}
  {\bibfnamefont {C.~H.}\ \bibnamefont {Back}}, \ and\ \bibinfo {author}
  {\bibfnamefont {B.~J.}\ \bibnamefont {Hickey}},\ }\href
  {https://link.aps.org/doi/10.1103/PhysRevB.95.064409} {\bibfield  {journal}
  {\bibinfo  {journal} {Physical Review B}\ }\textbf {\bibinfo {volume} {95}},\
  \bibinfo {pages} {064409} (\bibinfo {year} {2017})}\BibitemShut {NoStop}%
\bibitem [{\citenamefont {Park}\ \emph {et~al.}(2002)\citenamefont {Park},
  \citenamefont {Eames}, \citenamefont {Engebretson}, \citenamefont
  {Berezovsky},\ and\ \citenamefont {Crowell}}]{Park02}%
  \BibitemOpen
  \bibfield  {author} {\bibinfo {author} {\bibfnamefont {J.~P.}\ \bibnamefont
  {Park}}, \bibinfo {author} {\bibfnamefont {P.}~\bibnamefont {Eames}},
  \bibinfo {author} {\bibfnamefont {D.~M.}\ \bibnamefont {Engebretson}},
  \bibinfo {author} {\bibfnamefont {J.}~\bibnamefont {Berezovsky}}, \ and\
  \bibinfo {author} {\bibfnamefont {P.~A.}\ \bibnamefont {Crowell}},\ }\href
  {\doibase 10.1103/PhysRevLett.89.277201} {\bibfield  {journal} {\bibinfo
  {journal} {Physical Review Letters}\ }\textbf {\bibinfo {volume} {89}},\
  \bibinfo {pages} {277201} (\bibinfo {year} {2002})}\BibitemShut {NoStop}%
\bibitem [{\citenamefont {Collet}\ \emph {et~al.}(2017)\citenamefont {Collet},
  \citenamefont {Gladii}, \citenamefont {Evelt}, \citenamefont {Bessonov},
  \citenamefont {Soumah}, \citenamefont {Bortolotti}, \citenamefont
  {Demokritov}, \citenamefont {Henry}, \citenamefont {Cros}, \citenamefont
  {Bailleul}, \citenamefont {Demidov},\ and\ \citenamefont {Anane}}]{Collet17}%
  \BibitemOpen
  \bibfield  {author} {\bibinfo {author} {\bibfnamefont {M.}~\bibnamefont
  {Collet}}, \bibinfo {author} {\bibfnamefont {O.}~\bibnamefont {Gladii}},
  \bibinfo {author} {\bibfnamefont {M.}~\bibnamefont {Evelt}}, \bibinfo
  {author} {\bibfnamefont {V.}~\bibnamefont {Bessonov}}, \bibinfo {author}
  {\bibfnamefont {L.}~\bibnamefont {Soumah}}, \bibinfo {author} {\bibfnamefont
  {P.}~\bibnamefont {Bortolotti}}, \bibinfo {author} {\bibfnamefont {S.~O.}\
  \bibnamefont {Demokritov}}, \bibinfo {author} {\bibfnamefont
  {Y.}~\bibnamefont {Henry}}, \bibinfo {author} {\bibfnamefont
  {V.}~\bibnamefont {Cros}}, \bibinfo {author} {\bibfnamefont {M.}~\bibnamefont
  {Bailleul}}, \bibinfo {author} {\bibfnamefont {V.~E.}\ \bibnamefont
  {Demidov}}, \ and\ \bibinfo {author} {\bibfnamefont {A.}~\bibnamefont
  {Anane}},\ }\href {\doibase 10.1063/1.4976708} {\bibfield  {journal}
  {\bibinfo  {journal} {Applied Physics Letters}\ }\textbf {\bibinfo {volume}
  {110}},\ \bibinfo {pages} {092408} (\bibinfo {year} {2017})}\BibitemShut
  {NoStop}%
\bibitem [{\citenamefont {Sebastian}\ \emph {et~al.}(2015)\citenamefont
  {Sebastian}, \citenamefont {Schultheiss}, \citenamefont {Obry}, \citenamefont
  {Hillebrands},\ and\ \citenamefont {Schultheiss}}]{Sebastian15}%
  \BibitemOpen
  \bibfield  {author} {\bibinfo {author} {\bibfnamefont {T.}~\bibnamefont
  {Sebastian}}, \bibinfo {author} {\bibfnamefont {K.}~\bibnamefont
  {Schultheiss}}, \bibinfo {author} {\bibfnamefont {B.}~\bibnamefont {Obry}},
  \bibinfo {author} {\bibfnamefont {B.}~\bibnamefont {Hillebrands}}, \ and\
  \bibinfo {author} {\bibfnamefont {H.}~\bibnamefont {Schultheiss}},\ }\href
  {\doibase 10.3389/fphy.2015.00035} {\bibfield  {journal} {\bibinfo  {journal}
  {Frontiers in Physics}\ }\textbf {\bibinfo {volume} {3}},\ \bibinfo {pages}
  {35} (\bibinfo {year} {2015})}\BibitemShut {NoStop}%
\bibitem [{\citenamefont {Yu}\ \emph {et~al.}(2016)\citenamefont {Yu},
  \citenamefont {d’ Allivy~Kelly}, \citenamefont {Cros}, \citenamefont
  {Bernard}, \citenamefont {Bortolotti}, \citenamefont {Anane}, \citenamefont
  {Brandl}, \citenamefont {Heimbach},\ and\ \citenamefont {Grundler}}]{Yu16}%
  \BibitemOpen
  \bibfield  {author} {\bibinfo {author} {\bibfnamefont {H.}~\bibnamefont
  {Yu}}, \bibinfo {author} {\bibfnamefont {O.}~\bibnamefont {d’
  Allivy~Kelly}}, \bibinfo {author} {\bibfnamefont {V.}~\bibnamefont {Cros}},
  \bibinfo {author} {\bibfnamefont {R.}~\bibnamefont {Bernard}}, \bibinfo
  {author} {\bibfnamefont {P.}~\bibnamefont {Bortolotti}}, \bibinfo {author}
  {\bibfnamefont {A.}~\bibnamefont {Anane}}, \bibinfo {author} {\bibfnamefont
  {F.}~\bibnamefont {Brandl}}, \bibinfo {author} {\bibfnamefont
  {F.}~\bibnamefont {Heimbach}}, \ and\ \bibinfo {author} {\bibfnamefont
  {D.}~\bibnamefont {Grundler}},\ }\href {\doibase 10.1038/ncomms11255
  https://www.nature.com/articles/ncomms11255#supplementary-information} {\
  \textbf {\bibinfo {volume} {7}},\ \bibinfo {pages} {11255} (\bibinfo {year}
  {2016})}\BibitemShut {NoStop}%
\bibitem [{\citenamefont {Bailleul}\ \emph {et~al.}(2003)\citenamefont
  {Bailleul}, \citenamefont {Olligs},\ and\ \citenamefont
  {Fermon}}]{Bailleul03}%
  \BibitemOpen
  \bibfield  {author} {\bibinfo {author} {\bibfnamefont {M.}~\bibnamefont
  {Bailleul}}, \bibinfo {author} {\bibfnamefont {D.}~\bibnamefont {Olligs}}, \
  and\ \bibinfo {author} {\bibfnamefont {C.}~\bibnamefont {Fermon}},\ }\href
  {\doibase 10.1063/1.1597745} {\bibfield  {journal} {\bibinfo  {journal}
  {Applied Physics Letters}\ }\textbf {\bibinfo {volume} {83}},\ \bibinfo
  {pages} {972} (\bibinfo {year} {2003})}\BibitemShut {NoStop}%
\bibitem [{\citenamefont {de~Loubens}\ \emph {et~al.}(2007)\citenamefont
  {de~Loubens}, \citenamefont {Naletov}, \citenamefont {Klein}, \citenamefont
  {Youssef}, \citenamefont {Boust},\ and\ \citenamefont
  {Vukadinovic}}]{deLoubens07}%
  \BibitemOpen
  \bibfield  {author} {\bibinfo {author} {\bibfnamefont {G.}~\bibnamefont
  {de~Loubens}}, \bibinfo {author} {\bibfnamefont {V.~V.}\ \bibnamefont
  {Naletov}}, \bibinfo {author} {\bibfnamefont {O.}~\bibnamefont {Klein}},
  \bibinfo {author} {\bibfnamefont {J.~B.}\ \bibnamefont {Youssef}}, \bibinfo
  {author} {\bibfnamefont {F.}~\bibnamefont {Boust}}, \ and\ \bibinfo {author}
  {\bibfnamefont {N.}~\bibnamefont {Vukadinovic}},\ }\href {\doibase
  10.1103/PhysRevLett.98.127601} {\bibfield  {journal} {\bibinfo  {journal}
  {Physical Review Letters}\ }\textbf {\bibinfo {volume} {98}},\ \bibinfo
  {pages} {127601} (\bibinfo {year} {2007})}\BibitemShut {NoStop}%
\bibitem [{\citenamefont {Van~Waeyenberge}\ \emph {et~al.}(2006)\citenamefont
  {Van~Waeyenberge}, \citenamefont {Puzic}, \citenamefont {Stoll},
  \citenamefont {Chou}, \citenamefont {Tyliszczak}, \citenamefont {Hertel},
  \citenamefont {F\"ahnle}, \citenamefont {Br\"uckl}, \citenamefont {Rott},
  \citenamefont {Reiss}, \citenamefont {Neudecker}, \citenamefont {Weiss},
  \citenamefont {Back},\ and\ \citenamefont {Sch\"utz}}]{VanWaey06}%
  \BibitemOpen
  \bibfield  {author} {\bibinfo {author} {\bibfnamefont {B.}~\bibnamefont
  {Van~Waeyenberge}}, \bibinfo {author} {\bibfnamefont {A.}~\bibnamefont
  {Puzic}}, \bibinfo {author} {\bibfnamefont {H.}~\bibnamefont {Stoll}},
  \bibinfo {author} {\bibfnamefont {K.~W.}\ \bibnamefont {Chou}}, \bibinfo
  {author} {\bibfnamefont {T.}~\bibnamefont {Tyliszczak}}, \bibinfo {author}
  {\bibfnamefont {R.}~\bibnamefont {Hertel}}, \bibinfo {author} {\bibfnamefont
  {M.}~\bibnamefont {F\"ahnle}}, \bibinfo {author} {\bibfnamefont
  {H.}~\bibnamefont {Br\"uckl}}, \bibinfo {author} {\bibfnamefont
  {K.}~\bibnamefont {Rott}}, \bibinfo {author} {\bibfnamefont {G.}~\bibnamefont
  {Reiss}}, \bibinfo {author} {\bibfnamefont {I.}~\bibnamefont {Neudecker}},
  \bibinfo {author} {\bibfnamefont {D.}~\bibnamefont {Weiss}}, \bibinfo
  {author} {\bibfnamefont {C.~H.}\ \bibnamefont {Back}}, \ and\ \bibinfo
  {author} {\bibfnamefont {G.}~\bibnamefont {Sch\"utz}},\ }\href {\doibase
  10.1038/nature05240
  https://www.nature.com/articles/nature05240#supplementary-information}
  {\bibfield  {journal} {\bibinfo  {journal} {Nature}\ }\textbf {\bibinfo
  {volume} {444}},\ \bibinfo {pages} {461} (\bibinfo {year}
  {2006})}\BibitemShut {NoStop}%
\bibitem [{\citenamefont {Nolle}\ \emph {et~al.}(2012)\citenamefont {Nolle},
  \citenamefont {Weigand}, \citenamefont {Audehm}, \citenamefont {Goering},
  \citenamefont {Wiesemann}, \citenamefont {Wolter}, \citenamefont {Nolle},\
  and\ \citenamefont {Sch\"utz}}]{Nolle12}%
  \BibitemOpen
  \bibfield  {author} {\bibinfo {author} {\bibfnamefont {D.}~\bibnamefont
  {Nolle}}, \bibinfo {author} {\bibfnamefont {M.}~\bibnamefont {Weigand}},
  \bibinfo {author} {\bibfnamefont {P.}~\bibnamefont {Audehm}}, \bibinfo
  {author} {\bibfnamefont {E.}~\bibnamefont {Goering}}, \bibinfo {author}
  {\bibfnamefont {U.}~\bibnamefont {Wiesemann}}, \bibinfo {author}
  {\bibfnamefont {C.}~\bibnamefont {Wolter}}, \bibinfo {author} {\bibfnamefont
  {E.}~\bibnamefont {Nolle}}, \ and\ \bibinfo {author} {\bibfnamefont
  {G.}~\bibnamefont {Sch\"utz}},\ }\href {\doibase 10.1063/1.4707747}
  {\bibfield  {journal} {\bibinfo  {journal} {Review of Scientific
  Instruments}\ }\textbf {\bibinfo {volume} {83}},\ \bibinfo {pages} {046112}
  (\bibinfo {year} {2012})}\BibitemShut {NoStop}%
\bibitem [{\citenamefont {Noske}\ \emph {et~al.}(2014)\citenamefont {Noske},
  \citenamefont {Gangwar}, \citenamefont {Stoll}, \citenamefont {Kammerer},
  \citenamefont {Sproll}, \citenamefont {Dieterle}, \citenamefont {Weigand},
  \citenamefont {F\"ahnle}, \citenamefont {Woltersdorf}, \citenamefont {Back},\
  and\ \citenamefont {Sch\"utz}}]{Noske14}%
  \BibitemOpen
  \bibfield  {author} {\bibinfo {author} {\bibfnamefont {M.}~\bibnamefont
  {Noske}}, \bibinfo {author} {\bibfnamefont {A.}~\bibnamefont {Gangwar}},
  \bibinfo {author} {\bibfnamefont {H.}~\bibnamefont {Stoll}}, \bibinfo
  {author} {\bibfnamefont {M.}~\bibnamefont {Kammerer}}, \bibinfo {author}
  {\bibfnamefont {M.}~\bibnamefont {Sproll}}, \bibinfo {author} {\bibfnamefont
  {G.}~\bibnamefont {Dieterle}}, \bibinfo {author} {\bibfnamefont
  {M.}~\bibnamefont {Weigand}}, \bibinfo {author} {\bibfnamefont
  {M.}~\bibnamefont {F\"ahnle}}, \bibinfo {author} {\bibfnamefont
  {G.}~\bibnamefont {Woltersdorf}}, \bibinfo {author} {\bibfnamefont {C.~H.}\
  \bibnamefont {Back}}, \ and\ \bibinfo {author} {\bibfnamefont
  {G.}~\bibnamefont {Sch\"utz}},\ }\href {\doibase 10.1103/PhysRevB.90.104415}
  {\bibfield  {journal} {\bibinfo  {journal} {Physical Review B}\ }\textbf
  {\bibinfo {volume} {90}},\ \bibinfo {pages} {104415} (\bibinfo {year}
  {2014})}\BibitemShut {NoStop}%
\bibitem [{\citenamefont {Dieterle}\ \emph {et~al.}(2017)\citenamefont
  {Dieterle}, \citenamefont {F\"orster}, \citenamefont {Stoll}, \citenamefont
  {Semisalova}, \citenamefont {F\"ahnle}, \citenamefont {Bykova}, \citenamefont
  {Bozhko}, \citenamefont {Musiienko-Shmarova}, \citenamefont {Tiberkevich},
  \citenamefont {Slavin}, \citenamefont {Back}, \citenamefont {Raabe},
  \citenamefont {Sch\"utz},\ and\ \citenamefont {Wintz}}]{Dieterle17}%
  \BibitemOpen
  \bibfield  {author} {\bibinfo {author} {\bibfnamefont {G.}~\bibnamefont
  {Dieterle}}, \bibinfo {author} {\bibfnamefont {J.}~\bibnamefont {F\"orster}},
  \bibinfo {author} {\bibfnamefont {H.}~\bibnamefont {Stoll}}, \bibinfo
  {author} {\bibfnamefont {A.~S.}\ \bibnamefont {Semisalova}}, \bibinfo
  {author} {\bibfnamefont {M.}~\bibnamefont {F\"ahnle}}, \bibinfo {author}
  {\bibfnamefont {I.}~\bibnamefont {Bykova}}, \bibinfo {author} {\bibfnamefont
  {D.~A.}\ \bibnamefont {Bozhko}}, \bibinfo {author} {\bibfnamefont {H.~Y.}\
  \bibnamefont {Musiienko-Shmarova}}, \bibinfo {author} {\bibfnamefont
  {V.}~\bibnamefont {Tiberkevich}}, \bibinfo {author} {\bibfnamefont {A.~N.}\
  \bibnamefont {Slavin}}, \bibinfo {author} {\bibfnamefont {C.~H.}\
  \bibnamefont {Back}}, \bibinfo {author} {\bibfnamefont {J.}~\bibnamefont
  {Raabe}}, \bibinfo {author} {\bibfnamefont {G.}~\bibnamefont {Sch\"utz}}, \
  and\ \bibinfo {author} {\bibfnamefont {S.}~\bibnamefont {Wintz}},\ }\href
  {http://adsabs.harvard.edu/abs/2017arXiv171200681D} {\bibfield  {journal}
  {\bibinfo  {journal} {eprint arXiv:1712.00681 [cond-mat.mes-hall]}\ }
  (\bibinfo {year} {2017})}\BibitemShut {NoStop}%
\bibitem [{\citenamefont {Wintz}\ \emph {et~al.}(2016)\citenamefont {Wintz},
  \citenamefont {Tiberkevich}, \citenamefont {Weigand}, \citenamefont {Raabe},
  \citenamefont {Lindner}, \citenamefont {Erbe}, \citenamefont {Slavin},\ and\
  \citenamefont {Fassbender}}]{Wintz16}%
  \BibitemOpen
  \bibfield  {author} {\bibinfo {author} {\bibfnamefont {S.}~\bibnamefont
  {Wintz}}, \bibinfo {author} {\bibfnamefont {V.}~\bibnamefont {Tiberkevich}},
  \bibinfo {author} {\bibfnamefont {M.}~\bibnamefont {Weigand}}, \bibinfo
  {author} {\bibfnamefont {J.}~\bibnamefont {Raabe}}, \bibinfo {author}
  {\bibfnamefont {J.}~\bibnamefont {Lindner}}, \bibinfo {author} {\bibfnamefont
  {A.}~\bibnamefont {Erbe}}, \bibinfo {author} {\bibfnamefont {A.}~\bibnamefont
  {Slavin}}, \ and\ \bibinfo {author} {\bibfnamefont {J.}~\bibnamefont
  {Fassbender}},\ }\href {\doibase 10.1038/nnano.2016.117
  https://www.nature.com/articles/nnano.2016.117#supplementary-information}
  {\bibfield  {journal} {\bibinfo  {journal} {Nature Nanotechnology}\ }\textbf
  {\bibinfo {volume} {11}},\ \bibinfo {pages} {948} (\bibinfo {year}
  {2016})}\BibitemShut {NoStop}%
\bibitem [{\citenamefont {Gr\"afe}\ \emph {et~al.}(2017)\citenamefont
  {Gr\"afe}, \citenamefont {Decker}, \citenamefont {Keskinbora}, \citenamefont
  {Noske}, \citenamefont {Gawronski}, \citenamefont {Stoll}, \citenamefont
  {Back}, \citenamefont {Goering},\ and\ \citenamefont {Sch\"utz}}]{Graefe17}%
  \BibitemOpen
  \bibfield  {author} {\bibinfo {author} {\bibfnamefont {J.}~\bibnamefont
  {Gr\"afe}}, \bibinfo {author} {\bibfnamefont {M.}~\bibnamefont {Decker}},
  \bibinfo {author} {\bibfnamefont {K.}~\bibnamefont {Keskinbora}}, \bibinfo
  {author} {\bibfnamefont {M.}~\bibnamefont {Noske}}, \bibinfo {author}
  {\bibfnamefont {P.}~\bibnamefont {Gawronski}}, \bibinfo {author}
  {\bibfnamefont {H.}~\bibnamefont {Stoll}}, \bibinfo {author} {\bibfnamefont
  {C.}~\bibnamefont {Back}}, \bibinfo {author} {\bibfnamefont {E.}~\bibnamefont
  {Goering}}, \ and\ \bibinfo {author} {\bibfnamefont {G.}~\bibnamefont
  {Sch\"utz}},\ }\href {https://arxiv.org/abs/1707.03664v1} {\bibfield
  {journal} {\bibinfo  {journal} {eprint arXiv:1707.03664 [cond-mat.mes-hall]}\
  } (\bibinfo {year} {2017})}\BibitemShut {NoStop}%
\bibitem [{\citenamefont {Kammerer}\ \emph {et~al.}(2011)\citenamefont
  {Kammerer}, \citenamefont {Weigand}, \citenamefont {Curcic}, \citenamefont
  {Noske}, \citenamefont {Sproll}, \citenamefont {Vansteenkiste}, \citenamefont
  {Van~Waeyenberge}, \citenamefont {Stoll}, \citenamefont {Woltersdorf},
  \citenamefont {Back},\ and\ \citenamefont {Schuetz}}]{Kammerer11}%
  \BibitemOpen
  \bibfield  {author} {\bibinfo {author} {\bibfnamefont {M.}~\bibnamefont
  {Kammerer}}, \bibinfo {author} {\bibfnamefont {M.}~\bibnamefont {Weigand}},
  \bibinfo {author} {\bibfnamefont {M.}~\bibnamefont {Curcic}}, \bibinfo
  {author} {\bibfnamefont {M.}~\bibnamefont {Noske}}, \bibinfo {author}
  {\bibfnamefont {M.}~\bibnamefont {Sproll}}, \bibinfo {author} {\bibfnamefont
  {A.}~\bibnamefont {Vansteenkiste}}, \bibinfo {author} {\bibfnamefont
  {B.}~\bibnamefont {Van~Waeyenberge}}, \bibinfo {author} {\bibfnamefont
  {H.}~\bibnamefont {Stoll}}, \bibinfo {author} {\bibfnamefont
  {G.}~\bibnamefont {Woltersdorf}}, \bibinfo {author} {\bibfnamefont {C.~H.}\
  \bibnamefont {Back}}, \ and\ \bibinfo {author} {\bibfnamefont
  {G.}~\bibnamefont {Schuetz}},\ }\href {\doibase 10.1038/ncomms1277}
  {\bibfield  {journal} {\bibinfo  {journal} {Nature Communications}\ }\textbf
  {\bibinfo {volume} {2}},\ \bibinfo {pages} {279} (\bibinfo {year}
  {2011})}\BibitemShut {NoStop}%
\bibitem [{\citenamefont {Gro{\ss}}\ \emph {et~al.}(2019)\citenamefont
  {Gro{\ss}}, \citenamefont {Tr\"ager}, \citenamefont {F\"orster},
  \citenamefont {Weigand}, \citenamefont {Sch\"utz},\ and\ \citenamefont
  {Gr\"afe}}]{Gross19}%
  \BibitemOpen
  \bibfield  {author} {\bibinfo {author} {\bibfnamefont {F.}~\bibnamefont
  {Gro{\ss}}}, \bibinfo {author} {\bibfnamefont {N.}~\bibnamefont {Tr\"ager}},
  \bibinfo {author} {\bibfnamefont {J.}~\bibnamefont {F\"orster}}, \bibinfo
  {author} {\bibfnamefont {M.}~\bibnamefont {Weigand}}, \bibinfo {author}
  {\bibfnamefont {G.}~\bibnamefont {Sch\"utz}}, \ and\ \bibinfo {author}
  {\bibfnamefont {J.}~\bibnamefont {Gr\"afe}},\ }\href {\doibase
  10.1063/1.5074169} {\bibfield  {journal} {\bibinfo  {journal} {Applied
  Physics Letters}\ }\textbf {\bibinfo {volume} {114}},\ \bibinfo {pages}
  {012406} (\bibinfo {year} {2019})}\BibitemShut {NoStop}%
\bibitem [{\citenamefont {Simmendinger}\ \emph {et~al.}(2018)\citenamefont
  {Simmendinger}, \citenamefont {Ruoss}, \citenamefont {Stahl}, \citenamefont
  {Weigand}, \citenamefont {Gr\"afe}, \citenamefont {Sch\"utz},\ and\
  \citenamefont {Albrecht}}]{Simmendinger18}%
  \BibitemOpen
  \bibfield  {author} {\bibinfo {author} {\bibfnamefont {J.}~\bibnamefont
  {Simmendinger}}, \bibinfo {author} {\bibfnamefont {S.}~\bibnamefont {Ruoss}},
  \bibinfo {author} {\bibfnamefont {C.}~\bibnamefont {Stahl}}, \bibinfo
  {author} {\bibfnamefont {M.}~\bibnamefont {Weigand}}, \bibinfo {author}
  {\bibfnamefont {J.}~\bibnamefont {Gr\"afe}}, \bibinfo {author} {\bibfnamefont
  {G.}~\bibnamefont {Sch\"utz}}, \ and\ \bibinfo {author} {\bibfnamefont
  {J.}~\bibnamefont {Albrecht}},\ }\href {\doibase 10.1103/PhysRevB.97.134515}
  {\bibfield  {journal} {\bibinfo  {journal} {Physical Review B}\ }\textbf
  {\bibinfo {volume} {97}},\ \bibinfo {pages} {134515} (\bibinfo {year}
  {2018})}\BibitemShut {NoStop}%
\bibitem [{\citenamefont {Fohler}\ \emph {et~al.}(2017)\citenamefont {Fohler},
  \citenamefont {Fr\"ommel}, \citenamefont {Schneider}, \citenamefont {Pfau},
  \citenamefont {G\"unther}, \citenamefont {Hennecke}, \citenamefont {Guehrs},
  \citenamefont {Shemilt}, \citenamefont {Mishra}, \citenamefont {Berger},
  \citenamefont {Selve}, \citenamefont {Mitin}, \citenamefont {Albrecht},\ and\
  \citenamefont {Eisebitt}}]{Eisebitt17}%
  \BibitemOpen
  \bibfield  {author} {\bibinfo {author} {\bibfnamefont {M.}~\bibnamefont
  {Fohler}}, \bibinfo {author} {\bibfnamefont {S.}~\bibnamefont {Fr\"ommel}},
  \bibinfo {author} {\bibfnamefont {M.}~\bibnamefont {Schneider}}, \bibinfo
  {author} {\bibfnamefont {B.}~\bibnamefont {Pfau}}, \bibinfo {author}
  {\bibfnamefont {C.~M.}\ \bibnamefont {G\"unther}}, \bibinfo {author}
  {\bibfnamefont {M.}~\bibnamefont {Hennecke}}, \bibinfo {author}
  {\bibfnamefont {E.}~\bibnamefont {Guehrs}}, \bibinfo {author} {\bibfnamefont
  {L.}~\bibnamefont {Shemilt}}, \bibinfo {author} {\bibfnamefont
  {D.}~\bibnamefont {Mishra}}, \bibinfo {author} {\bibfnamefont
  {D.}~\bibnamefont {Berger}}, \bibinfo {author} {\bibfnamefont
  {S.}~\bibnamefont {Selve}}, \bibinfo {author} {\bibfnamefont
  {D.}~\bibnamefont {Mitin}}, \bibinfo {author} {\bibfnamefont
  {M.}~\bibnamefont {Albrecht}}, \ and\ \bibinfo {author} {\bibfnamefont
  {S.}~\bibnamefont {Eisebitt}},\ }\href {\doibase 10.1063/1.5006522}
  {\bibfield  {journal} {\bibinfo  {journal} {Review of Scientific
  Instruments}\ }\textbf {\bibinfo {volume} {88}},\ \bibinfo {pages} {103701}
  (\bibinfo {year} {2017})}\BibitemShut {NoStop}%
\bibitem [{\citenamefont {Dubs}\ \emph {et~al.}(2017)\citenamefont {Dubs},
  \citenamefont {Surzhenko}, \citenamefont {Linke}, \citenamefont {Danilewsky},
  \citenamefont {Br\"uckner},\ and\ \citenamefont {Dellith}}]{Dubs17}%
  \BibitemOpen
  \bibfield  {author} {\bibinfo {author} {\bibfnamefont {C.}~\bibnamefont
  {Dubs}}, \bibinfo {author} {\bibfnamefont {O.}~\bibnamefont {Surzhenko}},
  \bibinfo {author} {\bibfnamefont {R.}~\bibnamefont {Linke}}, \bibinfo
  {author} {\bibfnamefont {A.}~\bibnamefont {Danilewsky}}, \bibinfo {author}
  {\bibfnamefont {U.}~\bibnamefont {Br\"uckner}}, \ and\ \bibinfo {author}
  {\bibfnamefont {J.}~\bibnamefont {Dellith}},\ }\href
  {http://stacks.iop.org/0022-3727/50/i=20/a=204005} {\bibfield  {journal}
  {\bibinfo  {journal} {Journal of Physics D: Applied Physics}\ }\textbf
  {\bibinfo {volume} {50}},\ \bibinfo {pages} {204005} (\bibinfo {year}
  {2017})}\BibitemShut {NoStop}%
\bibitem [{\citenamefont {Pirro}\ \emph {et~al.}(2014)\citenamefont {Pirro},
  \citenamefont {Br\"acher}, \citenamefont {Chumak}, \citenamefont {L\"agel},
  \citenamefont {Dubs}, \citenamefont {Surzhenko}, \citenamefont {G\"ornert},
  \citenamefont {Leven},\ and\ \citenamefont {Hillebrands}}]{Pirro14}%
  \BibitemOpen
  \bibfield  {author} {\bibinfo {author} {\bibfnamefont {P.}~\bibnamefont
  {Pirro}}, \bibinfo {author} {\bibfnamefont {T.}~\bibnamefont {Br\"acher}},
  \bibinfo {author} {\bibfnamefont {A.~V.}\ \bibnamefont {Chumak}}, \bibinfo
  {author} {\bibfnamefont {B.}~\bibnamefont {L\"agel}}, \bibinfo {author}
  {\bibfnamefont {C.}~\bibnamefont {Dubs}}, \bibinfo {author} {\bibfnamefont
  {O.}~\bibnamefont {Surzhenko}}, \bibinfo {author} {\bibfnamefont
  {P.}~\bibnamefont {G\"ornert}}, \bibinfo {author} {\bibfnamefont
  {B.}~\bibnamefont {Leven}}, \ and\ \bibinfo {author} {\bibfnamefont
  {B.}~\bibnamefont {Hillebrands}},\ }\href {\doibase 10.1063/1.4861343}
  {\bibfield  {journal} {\bibinfo  {journal} {Applied Physics Letters}\
  }\textbf {\bibinfo {volume} {104}},\ \bibinfo {pages} {012402} (\bibinfo
  {year} {2014})}\BibitemShut {NoStop}%
\bibitem [{\citenamefont {Krysztofik}\ \emph {et~al.}(2017)\citenamefont
  {Krysztofik}, \citenamefont {Coy}, \citenamefont {Ku\'swik}, \citenamefont
  {Za\l{}\k{e}ski}, \citenamefont {G\l{}owi\'nski},\ and\ \citenamefont
  {Dubowik}}]{Krysztofik17}%
  \BibitemOpen
  \bibfield  {author} {\bibinfo {author} {\bibfnamefont {A.}~\bibnamefont
  {Krysztofik}}, \bibinfo {author} {\bibfnamefont {L.~E.}\ \bibnamefont {Coy}},
  \bibinfo {author} {\bibfnamefont {P.}~\bibnamefont {Ku\'swik}}, \bibinfo
  {author} {\bibfnamefont {K.}~\bibnamefont {Za\l{}\k{e}ski}}, \bibinfo
  {author} {\bibfnamefont {H.}~\bibnamefont {G\l{}owi\'nski}}, \ and\ \bibinfo
  {author} {\bibfnamefont {J.}~\bibnamefont {Dubowik}},\ }\href {\doibase
  10.1063/1.5002004} {\bibfield  {journal} {\bibinfo  {journal} {Applied
  Physics Letters}\ }\textbf {\bibinfo {volume} {111}},\ \bibinfo {pages}
  {192404} (\bibinfo {year} {2017})}\BibitemShut {NoStop}%
\bibitem [{\citenamefont {{ Ernst Ruska-Centre for Microscopy and Spectroscopy
  with Electrons (ER-C) et al.}}(2016)}]{ErnstRuska16}%
  \BibitemOpen
  \bibfield  {author} {\bibinfo {author} {\bibnamefont {{ Ernst Ruska-Centre
  for Microscopy and Spectroscopy with Electrons (ER-C) et al.}}},\ }\href@noop
  {} {\bibfield  {journal} {\bibinfo  {journal} {Journal of large-scale
  research facilities}\ }\textbf {\bibinfo {volume} {2}},\ \bibinfo {pages}
  {A42} (\bibinfo {year} {2016})}\BibitemShut {NoStop}%
\bibitem [{\citenamefont {Sch\"utz}\ \emph {et~al.}(1987)\citenamefont
  {Sch\"utz}, \citenamefont {Wagner}, \citenamefont {Wilhelm}, \citenamefont
  {Kienle}, \citenamefont {Zeller}, \citenamefont {Frahm},\ and\ \citenamefont
  {Materlik}}]{Schuetz87}%
  \BibitemOpen
  \bibfield  {author} {\bibinfo {author} {\bibfnamefont {G.}~\bibnamefont
  {Sch\"utz}}, \bibinfo {author} {\bibfnamefont {W.}~\bibnamefont {Wagner}},
  \bibinfo {author} {\bibfnamefont {W.}~\bibnamefont {Wilhelm}}, \bibinfo
  {author} {\bibfnamefont {P.}~\bibnamefont {Kienle}}, \bibinfo {author}
  {\bibfnamefont {R.}~\bibnamefont {Zeller}}, \bibinfo {author} {\bibfnamefont
  {R.}~\bibnamefont {Frahm}}, \ and\ \bibinfo {author} {\bibfnamefont
  {G.}~\bibnamefont {Materlik}},\ }\href {\doibase 10.1103/PhysRevLett.58.737}
  {\bibfield  {journal} {\bibinfo  {journal} {Physical Review Letters}\
  }\textbf {\bibinfo {volume} {58}},\ \bibinfo {pages} {737} (\bibinfo {year}
  {1987})}\BibitemShut {NoStop}%
\bibitem [{\citenamefont {Krichevtsov}\ \emph {et~al.}(2017)\citenamefont
  {Krichevtsov}, \citenamefont {Gastev}, \citenamefont {Suturin}, \citenamefont
  {Fedorov}, \citenamefont {Korovin}, \citenamefont {Bursian}, \citenamefont
  {Banshchikov}, \citenamefont {Volkov}, \citenamefont {Tabuchi},\ and\
  \citenamefont {Sokolov}}]{Krichevtsov17}%
  \BibitemOpen
  \bibfield  {author} {\bibinfo {author} {\bibfnamefont {B.~B.}\ \bibnamefont
  {Krichevtsov}}, \bibinfo {author} {\bibfnamefont {S.~V.}\ \bibnamefont
  {Gastev}}, \bibinfo {author} {\bibfnamefont {S.~M.}\ \bibnamefont {Suturin}},
  \bibinfo {author} {\bibfnamefont {V.~V.}\ \bibnamefont {Fedorov}}, \bibinfo
  {author} {\bibfnamefont {A.~M.}\ \bibnamefont {Korovin}}, \bibinfo {author}
  {\bibfnamefont {V.~E.}\ \bibnamefont {Bursian}}, \bibinfo {author}
  {\bibfnamefont {A.~G.}\ \bibnamefont {Banshchikov}}, \bibinfo {author}
  {\bibfnamefont {M.~P.}\ \bibnamefont {Volkov}}, \bibinfo {author}
  {\bibfnamefont {M.}~\bibnamefont {Tabuchi}}, \ and\ \bibinfo {author}
  {\bibfnamefont {N.~S.}\ \bibnamefont {Sokolov}},\ }\href {\doibase
  10.1080/14686996.2017.1316422} {\bibfield  {journal} {\bibinfo  {journal}
  {Science and Technology of Advanced Materials}\ }\textbf {\bibinfo {volume}
  {18}},\ \bibinfo {pages} {351} (\bibinfo {year} {2017})}\BibitemShut
  {NoStop}%
\bibitem [{\citenamefont {Oliphant}(2007)}]{Numpy}%
  \BibitemOpen
  \bibfield  {author} {\bibinfo {author} {\bibfnamefont {T.~E.}\ \bibnamefont
  {Oliphant}},\ }\href {\doibase 10.1109/MCSE.2007.58} {\bibfield  {journal}
  {\bibinfo  {journal} {Computing in Science \& Engineering}\ }\textbf
  {\bibinfo {volume} {9}},\ \bibinfo {pages} {10} (\bibinfo {year}
  {2007})}\BibitemShut {NoStop}%
\bibitem [{\citenamefont {Hunter}(2007)}]{Matplotlib}%
  \BibitemOpen
  \bibfield  {author} {\bibinfo {author} {\bibfnamefont {J.~D.}\ \bibnamefont
  {Hunter}},\ }\href {\doibase 10.1109/MCSE.2007.55} {\bibfield  {journal}
  {\bibinfo  {journal} {Computing In Science \& Engineering}\ }\textbf
  {\bibinfo {volume} {9}},\ \bibinfo {pages} {90} (\bibinfo {year}
  {2007})}\BibitemShut {NoStop}%
\bibitem [{\citenamefont {Jorzick}\ \emph {et~al.}(2002)\citenamefont
  {Jorzick}, \citenamefont {Demokritov}, \citenamefont {Hillebrands},
  \citenamefont {Bailleul}, \citenamefont {Fermon}, \citenamefont {Guslienko},
  \citenamefont {Slavin}, \citenamefont {Berkov},\ and\ \citenamefont
  {Gorn}}]{Jorzick02}%
  \BibitemOpen
  \bibfield  {author} {\bibinfo {author} {\bibfnamefont {J.}~\bibnamefont
  {Jorzick}}, \bibinfo {author} {\bibfnamefont {S.~O.}\ \bibnamefont
  {Demokritov}}, \bibinfo {author} {\bibfnamefont {B.}~\bibnamefont
  {Hillebrands}}, \bibinfo {author} {\bibfnamefont {M.}~\bibnamefont
  {Bailleul}}, \bibinfo {author} {\bibfnamefont {C.}~\bibnamefont {Fermon}},
  \bibinfo {author} {\bibfnamefont {K.~Y.}\ \bibnamefont {Guslienko}}, \bibinfo
  {author} {\bibfnamefont {A.~N.}\ \bibnamefont {Slavin}}, \bibinfo {author}
  {\bibfnamefont {D.~V.}\ \bibnamefont {Berkov}}, \ and\ \bibinfo {author}
  {\bibfnamefont {N.~L.}\ \bibnamefont {Gorn}},\ }\href {\doibase
  10.1103/PhysRevLett.88.047204} {\bibfield  {journal} {\bibinfo  {journal}
  {Physical Review Letters}\ }\textbf {\bibinfo {volume} {88}},\ \bibinfo
  {pages} {047204} (\bibinfo {year} {2002})}\BibitemShut {NoStop}%
\bibitem [{\citenamefont {Bayer}\ \emph {et~al.}(2003)\citenamefont {Bayer},
  \citenamefont {Demokritov}, \citenamefont {Hillebrands},\ and\ \citenamefont
  {Slavin}}]{Bayer03}%
  \BibitemOpen
  \bibfield  {author} {\bibinfo {author} {\bibfnamefont {C.}~\bibnamefont
  {Bayer}}, \bibinfo {author} {\bibfnamefont {S.~O.}\ \bibnamefont
  {Demokritov}}, \bibinfo {author} {\bibfnamefont {B.}~\bibnamefont
  {Hillebrands}}, \ and\ \bibinfo {author} {\bibfnamefont {A.~N.}\ \bibnamefont
  {Slavin}},\ }\href {\doibase 10.1063/1.1540734} {\bibfield  {journal}
  {\bibinfo  {journal} {Applied Physics Letters}\ }\textbf {\bibinfo {volume}
  {82}},\ \bibinfo {pages} {607} (\bibinfo {year} {2003})}\BibitemShut
  {NoStop}%
\bibitem [{\citenamefont {Puszkarski}\ \emph {et~al.}(2005)\citenamefont
  {Puszkarski}, \citenamefont {Krawczyk},\ and\ \citenamefont
  {Lévy}}]{Puszkar05}%
  \BibitemOpen
  \bibfield  {author} {\bibinfo {author} {\bibfnamefont {H.}~\bibnamefont
  {Puszkarski}}, \bibinfo {author} {\bibfnamefont {M.}~\bibnamefont
  {Krawczyk}}, \ and\ \bibinfo {author} {\bibfnamefont {J.~C.~S.}\ \bibnamefont
  {Lévy}},\ }\href {\doibase 10.1103/PhysRevB.71.014421} {\bibfield  {journal}
  {\bibinfo  {journal} {Physical Review B}\ }\textbf {\bibinfo {volume} {71}},\
  \bibinfo {pages} {014421} (\bibinfo {year} {2005})}\BibitemShut {NoStop}%
\bibitem [{Note1()}]{Note1}%
  \BibitemOpen
  \bibinfo {note} {Note that in the original Damon-Eshbach theory a non-uniform
  thickness profile is considered while the influence of exchange is neglected
  \cite {Damon65}.}\BibitemShut {Stop}%
\bibitem [{\citenamefont {Klingler}\ \emph
  {et~al.}(2015{\natexlab{b}})\citenamefont {Klingler}, \citenamefont {Chumak},
  \citenamefont {Mewes}, \citenamefont {Khodadadi}, \citenamefont {Mewes},
  \citenamefont {Dubs}, \citenamefont {Surzhenko}, \citenamefont
  {Hillebrands},\ and\ \citenamefont {Conca}}]{Klingler15}%
  \BibitemOpen
  \bibfield  {author} {\bibinfo {author} {\bibfnamefont {S.}~\bibnamefont
  {Klingler}}, \bibinfo {author} {\bibfnamefont {A.~V.}\ \bibnamefont
  {Chumak}}, \bibinfo {author} {\bibfnamefont {T.}~\bibnamefont {Mewes}},
  \bibinfo {author} {\bibfnamefont {B.}~\bibnamefont {Khodadadi}}, \bibinfo
  {author} {\bibfnamefont {C.}~\bibnamefont {Mewes}}, \bibinfo {author}
  {\bibfnamefont {C.}~\bibnamefont {Dubs}}, \bibinfo {author} {\bibfnamefont
  {O.}~\bibnamefont {Surzhenko}}, \bibinfo {author} {\bibfnamefont
  {B.}~\bibnamefont {Hillebrands}}, \ and\ \bibinfo {author} {\bibfnamefont
  {A.}~\bibnamefont {Conca}},\ }\href
  {http://stacks.iop.org/0022-3727/48/i=1/a=015001} {\bibfield  {journal}
  {\bibinfo  {journal} {Journal of Physics D: Applied Physics}\ }\textbf
  {\bibinfo {volume} {48}},\ \bibinfo {pages} {015001} (\bibinfo {year}
  {2015}{\natexlab{b}})}\BibitemShut {NoStop}%
\bibitem [{\citenamefont {Vansteenkiste}\ \emph {et~al.}(2014)\citenamefont
  {Vansteenkiste}, \citenamefont {Leliaert}, \citenamefont {Dvornik},
  \citenamefont {Helsen}, \citenamefont {Garcia-Sanchez},\ and\ \citenamefont
  {Waeyenberge}}]{Mumax}%
  \BibitemOpen
  \bibfield  {author} {\bibinfo {author} {\bibfnamefont {A.}~\bibnamefont
  {Vansteenkiste}}, \bibinfo {author} {\bibfnamefont {J.}~\bibnamefont
  {Leliaert}}, \bibinfo {author} {\bibfnamefont {M.}~\bibnamefont {Dvornik}},
  \bibinfo {author} {\bibfnamefont {M.}~\bibnamefont {Helsen}}, \bibinfo
  {author} {\bibfnamefont {F.}~\bibnamefont {Garcia-Sanchez}}, \ and\ \bibinfo
  {author} {\bibfnamefont {B.~V.}\ \bibnamefont {Waeyenberge}},\ }\href
  {\doibase 10.1063/1.4899186} {\bibfield  {journal} {\bibinfo  {journal} {AIP
  Advances}\ }\textbf {\bibinfo {volume} {4}},\ \bibinfo {pages} {107133}
  (\bibinfo {year} {2014})}\BibitemShut {NoStop}%
\end{thebibliography}%

\end{document}